\input{psfig}
\documentclass[12pt,preprint]{aastex}
\usepackage{maple2e}

\shorttitle{Measuring $\Omega_{\rm m}$ using clusters evolution}
\shortauthors{A. Del Popolo}

\begin{document}

%% LaTeX will automatically break titles if they run longer than
%% one line. However, you may use \\ to force a line break if
%% you desire.

\title{Improvements in the M-T relation and mass function and the
measured $\Omega_{\rm m}$ through clusters evolution}

%% Use \author, \affil, and the \and command to format
%% author and affiliation information.
%% Note that \email has replaced the old \authoremail command
%% from AASTeX v4.0. You can use \email to mark an email address
%% anywhere in the paper, not just in the front matter.
%% As in the title, you can use \\ to force line breaks.

\author{A. Del Popolo\altaffilmark{1,2}}

\affil{Dipartimento di Matematica, Universit\`{a} Statale di Bergamo,
  Piazza Rosate, 2 - I 24129 Bergamo, ITALY }
%% Notice that each of these authors has alternate affiliations, which
%% are identified by the \altaffilmark after each name.  Specify alternate
%% affiliation information with \altaffiltext, with one command per each
%% affiliation.

%\altaffiltext{2}{present address: Feza G\"ursey Institute, P.O. Box 6 \c Cengelk\"oy, Istanbul,
%     Turkey.}
\altaffiltext{2}{Bo$\breve{g}azi$\c{c}i University, Physics Department,
     80815 Bebek, Istanbul, Turkey
}

%% Mark off your abstract in the ``abstract'' environment. In the manuscript
%% style, abstract will output a Received/Accepted line after the
%% title and affiliation information. No date will appear since the author
%% does not have this information. The dates will be filled in by the
%% editorial office after submission.

\begin{abstract}
In this paper, I revisit the constraints obtained
by several authors (Reichart et al. 1999; Eke et al. 1998; Henry 2000)
on the estimated values of $\Omega_{\rm m}$, $n$ and $\sigma_8$ in the light of recent theoretical developments: 1) new theoretical mass functions (Sheth \& Tormen 1999, Sheth, Mo \& Tormen 1999, Del Popolo 2002b); 2) a more accurate mass-temperature relation, also determined for arbitrary $\Omega_{\rm m}$ and $\Omega_{\rm \Lambda}$ (Voit 2000, Pierpaoli et al. 2001, Del Popolo 2002a). Firstly, using the quoted improvements, I re-derive an expression for the X-ray Luminosity Function (XLF), similarly to Reichart et al. (1999), and then I get some constraints to $\Omega_{\rm m}$ and $n$, by using the {\it ROSAT} BCS and {\rm EMSS} samples and maximum-likelihood analysis. Then I re-derive the X-ray Temperature Function (XTF), similarly to Henry (2000) and Eke et al. (1999), re-obtaining the constraints on $\Omega_{\rm m}$, $n$, $\sigma_8$.
Both in the case of the XLF and XTF, the changes in the mass function and M-T relation produces an increase in $\Omega_{\rm m}$ of $ \simeq 20\%$ and similar results in $\sigma_8$ and $n$.
%Finally, I compare the normalization $\sigma_8$ as a function of $\Omega_{\rm m}$ obtained by Pierpaoli et al. (2001) using %the Sheth \& Tormen (1999) mass function and an improved M-T relation, with the constraints obtained from the modification %of Henry (2000) results.
\end{abstract}

%% Keywords should appear after the \end{abstract} command. The uncommented
%% example has been keyed in ApJ style. See the instructions to authors
%% for the journal to which you are submitting your paper to determine
%% what keyword punctuation is appropriate.

\keywords{cosmology: theory - large scale structure of universe - galaxies:
formation}

%% From the front matter, we move on to the body of the paper.
%% In the first two sections, notice the use of the natbib \citep
%% and \citet commands to identify citations.  The citations are
%% tied to the reference list via symbolic KEYs. The KEY corresponds
%% to the KEY in the \bibitem in the reference list below. We have
%% chosen the first three characters of the first author's name plus
%% the last two numeral of the year of publication as our KEY for
%% each reference.

\section{Introduction}

Galaxy clusters represents the virialization stage of exceptionally high peaks of initial density perturbations
on comoving scales of $\simeq 10 h^{-1} {\rm Mpc}$, and as such they provide useful cosmological probes. The evolution
in the abundance of clusters is strongly dependent on the cosmological density parameter, $\Omega_{\rm m}$ (Evrard 1989; Oukbir \& Blanchard 1992; Eke, Cole \& Frenk 1996; Donahue et al. 1998; Borgani et al. 1999): in a critical universe
with $\Omega_{\rm}=1$ , perturbation growth continues forever, while in a low-density universe ($\Omega_{\rm}<1$), growth significantly decelerates once $z \simeq \Omega_{\rm}^{-1}-1$. Although the quoted cosmological test is in principle very powerful, there are two main problems in practical applications. First, theoretical predictions provide the number density of clusters of a given mass, while the mass itself is never the directly observed quantity. Secondly, it is needed a cluster sample that spans a large $z$ baseline, and is based on model-independent selection criteria \footnote{So that the search volume and the number density associated with each cluster are uniquely identified}.
In this respect, X-ray observations provide a very efficient method to identify distant clusters down to a given X-ray flux limit, and hence within a known survey volume for each luminosity, $L_{\rm X}$. For this reason, most studies using clusters as cosmological probes are based on X-ray samples.
It is well known that clusters are strong X-ray emitters and so cluster evolution can be inferred from the study of X-ray properties of distant clusters. The amount of observational data concerning high-redshift cluster properties has increased in the past years. {\it EMSS} (Henry et al. 1992; Gioia \& Luppino 1994), {\it ASCA} (Donahue 1996; Henry 1997) and {\it ROSAT} (Ebeling et al. 1997; Rosati et al. 1998) measurements of the X-ray emitting intracluster plasma complements low-redshift studies carried out by Edge et al. (1990) and David et al. (1993). In addition, galaxy velocity dispersions for a well-defined sample of high-redshift clusters (Carlberg et al. 1996) are provided by the {\it CNOC} survey. The Press \& Schechter (1974) (hereafter PS) formalism has been heavily used to model the cluster population. The combination of the PS mass function and the X-ray cluster catalogs represents a unique opportunity to constraint cosmological parameters, (e.g. the mass density parameter, $\Omega_{\rm m}$). The PS approach has gained with time an increasing favor since this approach offers a number of advantages when compared with more traditional methods of measuring $\Omega_{\rm m}$.
Firstly, unlike methods that only probe $\Omega_{\rm m}$ over small spatial scales, the PS approach probes
$\Omega_{\rm m}$ over the scales of the X-ray cluster catalogs.
Secondly the PS approach seems to be relatively insensitive to the presence of a cosmological constant (Henry 1997; Eke et al. 1998; Viana \& Liddle 1999).
Thirdly, till some years ago, it was shown that numerical simulations reproduce the PS mass function quite accurately (Eke et al. 1996; Bryan \& Norman 1998). More recent studies has, however, shown some discrepancies between PS and simulations. Although the analytical framework of the PS model has been greatly
refined and extended (e.g., Lacey \& Cole 1993), it is well known that the PS mass function,
while qualitatively correct, disagrees with the results of
N-body simulations. In particular, the PS formula overestimates
the abundance of haloes near the characteristic mass
$M_{\ast}$ and underestimates the abundance in the high mass tail
(Efstathiou et al. 1988; Lacey \& Cole 1994;
Tozzi \& Governato 1998; Gross et al. 1998; Governato et al. 1999).
The quoted discrepancy is not surprising since the PS model, as any other analytical model,
should make several assumptions to get simple analytical predictions.

There are different methods to trace the evolution of the cluster number density: \\
a) The X-ray temperature function (XTF) has been presented for local (e.g. Henry \& Arnaud 1991) and distant clusters (Eke et al. 1998; Henry 2000). The mild evolution of the XTF has been interpreted as a strong indication for a low density universe ($0.2<\Omega_{\rm m}<0.6$). As described by Colafrancesco, Mazzotta \& Vittorio (1997), Viana \& Liddle 1999, and Blanchard et al. 2000, this conclusion could be weakened by uncertainties related to the limited amount of high-z data and to the lack of a homogeneous sample selection for local and distant clusters.\\
b) The evolution of the X-ray luminosity function (XLF). In this case, we need a relation between
the observed $L_{\rm x}$ and the cluster virial mass. Studies of the intra-cluster-medium (ICM) (Ponman, Cannon \& Navarro 1999) showed that non-gravitational heating and, possibly, radiative cooling significantly affects the $L_{\rm X}-M$ relation. At the same time, other studies (Borgani \& Guzzo 2001) showed the X-ray luminosity
to be a robust diagnostic of cluster masses. Furthermore, the most recent flux limited cluster samples contain a large ($\simeq 100$) number of objects, which are homogeneously identified over a broad resdshift baseline, out to $z \simeq 1.3$.

The results for $\Omega_{\rm m}$ obtained are several.\\
Kytayama \& Suto (1997) and Mathiesen \& Evrard (1998) analyzed the number counts from different X-ray flux-limited cluster surveys and found that the resulting constraints on $\Omega_{\rm m}$ are rather sensitive to the evolution of the mass luminosity function. Analyzing {\it EMSS}, Sadat, Blanchard
\& Oukbir (1998) and Reichart et al. (1999) found results consistent with $\Omega_{\rm m}=1$. A result consistent with $\Omega_{\rm m} \simeq 1$ was found by Blanchard \& Bartlett (1998), and Viana \& Liddle (1999) found that
$\Omega_{\rm m} \simeq 0.75$ with $\Omega_{\rm m}>0.3$ at the 90\% confidence level and $\Omega_{\rm m} \simeq 1$ still viable. Blanchard, Bartlett \& Sadat (1998) found almost identical results ($\Omega_{\rm m} \simeq 0.74$ with $0.3<\Omega_{\rm m}<1.2$ at the 95\% confidence level).
Eke et al. (1998) found $\Omega_{\rm m}=0.45 \pm 0.2$. It is interesting to note (as previously mentioned) that Viana \& Liddle (1999) used the same data set as Eke et al. (1998) and showed that uncertainties both in fitting local data and in the theoretical modeling could significantly change the final results: they found $\Omega_{\rm m} \simeq 0.75$ as a preferred value with a critical density model acceptable at $<90\%$ c.l.

Different results were obtained by Bahcall, Fan \& Cen (1997) ($\Omega_{\rm m}=0.3 \pm 0.1$), Fan, Bahcall \& Cen (1997) ($\Omega_{\rm m} \simeq 0.3 \pm 0.1$), Bahcall \& Fan (1998) ($\Omega_{\rm m}=0.2^{+0.3}_{-0.1}$). The previous example together with other not cited, shows
results span the entire range of acceptable solutions: $0.2 \leq \Omega_{\rm m} \leq 1$ (see Reichart et al. 1999).
%So summarizing, the value of $\Omega_{\rm m}$ obtained changes according to the method and data used and sometime the same %data can lead........\\

The reasons leading to the quoted discrepancies has been studied in several papers (Eke et al. 1998; Reichart et al. 1999; Donahue \& Voit 1999; Borgani et al. 2001).
According to Reichart (1999) unknown systematic effects may be plaguing great part of the quoted results. Systematic effects  entering the quoted analysis are: 1) The inadequate approximation given by the PS (e.g., Bryan \& Norman 1997). 2) Inadequacy in the structure formation as described by the spherical model leading to changes in the threshold parameter $\delta_{\rm c}$ (e.g., Governato et al. 1998). 3) Inadequacy in the M-T relation obtained from the virial theorem (see Voit \& Donahue 1998; Del Popolo 2002a). \footnote{Even if Reichart et al. (1999), did some estimation of the changes produced by the previous systematic effects, as stressed by the same authors a further investigation is needed taking into account the correct forms of the M-T relation and improved versions of the PS theory} 4) Effects of cowling flows. 5) Determination of the X-ray cluster catalog's selection function.

Voit \& Donahue (1999) point the attention on similar and different items to that stressed by Reichart et al. (1999):
1) Deviation from the Press-Schechter orthodoxy (similarly to Reichart et al. 1999). 2) Missing high redshift clusters in the data used (e.g., the {\it EMSS}). 3) Inadequacy in the M-T relation. 4) Evolution of the L-T relation.

Eke et al. (1998), in a detailed study of the systematic uncertainties in the determination of $\Omega_{\rm m}$ showed that
even the differences between the derived best-fitting parameters from the $\chi^2$ and the Maximum likelihood techniques has non-negligible contributions in the systematic uncertainties.
As stressed by Henry (2000),
the maximum likelihood fit only provides the best fit, it does not provide an assessment of whether that
fit is a good fit to the data. Marshall et al (1983) require that the two independent cumulants (see Eqs. 7, 8 of
Henry (2000)) both be uniform according to the two­-tailed Kolmogorov­-Smirnov test. They recommend rejecting the
model if the probability of either being uniform is less than 0.05.

Although the quoted uncertainties has been so far of minor importance with respect to the paucity of observational data, a breakthrough is needed in the quality of the theoretical framework if high-redshift clusters are to take part in the high-precision-era of observational cosmology.

Even if there have been several recent and detailed studies of
the cluster abundance (Bond \& Myers (1996), Eke,
Cole \& Frenk (1996), Viana \& Liddle (1996; 1999), Colafrancesco,
Mazzotta \& Vittorio (1997), Kitayama \& Suto
(1997), Eke et al. (1998), Pen (1998), Wang \& Steinhardt
(1998), Donahue \& Voit (1999) and Henry (2000))

The above discussion and
%however,
%even more recently,
recent
%there have been several
developments
in terms of both theory and observation which suggest it
would be useful to revisit this constraint:
1) the addition of ASCA
temperatures (Tanaka, Inoue \& Holt 1994)
means that there is now a well defined local temperature
function for clusters, with relatively small errors in temperature.
2) A number of large N-body simulations have
accurately determined the mass function of virialized haloes
(e.g. Governato et al. 1999), finding non-negligible deviations from the old
Press-Schechter (1974; hereafter PS) theory. 3) More ambitious hydrodynamical simulations of cluster
formation (e.g. Frenk et al. 1999, and references therein) and theoretical analysis
(Voit 2000, Del Popolo 2002a) have
resulted in improvements in the relationship between mass
and temperature and a better estimate of its scatter.

These reasons lead me to re-calculate the constraints on $\Omega_{\rm m}$, $n$ and $\sigma_8$, using the XLF and XTF.
The paper is organized as follows: in Sect. ~2, I re-calculate the XLF, as done by Reichart et al. (1999)
and obtained constraints for $\Omega_{\rm m}$ and $n$ using data from
{\it ROSAT} BCS and {\rm EMSS} samples.
%and I compare them with previous results and numerical simulations.
%In Sect. ~3, I do the same for the ``conditional" mass function.
In Sect. ~3, I re-calculate the XTF, as done by Henry (2000) and Eke et al. (1998) and obtained
constraints for $\Omega_{\rm m}$ and $n$ and $\sigma_8$.
Sect. ~4 is devoted to results and to conclusions.

\section{Constraints to $\Omega_{\rm m}$ and $n$ from the XLF}

In this section, similarly to Reichart et al. (1999), I derive an expression for the XLF (using now the mass function and M-T relation obtained in Del Popolo 2000a, Del Popolo 2000b, respectively) and
then I get some constraints to $\Omega_{\rm m}$ and $n$, by using the {\it ROSAT} BCS and {\rm EMSS} samples.

As previously quoted, although the analytical framework of the PS model has been greatly refined and extended, it is well known that the PS mass function, while qualitatively correct, disagrees with the results of
N-body simulations (Efstathiou et al. 1988; White, Efstathiou \& Frenk 1993; Lacey \& Cole 1994; Tozzi \& Governato 1998; Gross et al. 1998; Governato et al. 1999). The quoted discrepancy
is not surprising since the PS model, as any other analytical model, should make several assumptions to get simple
analytical predictions. As previously reported, the main assumptions that the PS model combines are the simple physics
of the spherical collapse model with the assumption that the initial fluctuations were Gaussian and small.
On average, initially denser regions collapse before less dense ones, which means that, at any given epoch,
there is a critical density, $\delta_c(z)$, which must be exceeded if collapse is to occur.
In the spherical collapse model, this critical density does not depend on the mass of the collapsed object. Taking account of the effects of asphericity and
tidal interaction with neighbors, Del Popolo \& Gambera (1998) and Sheth, Mo \& Tormen (2001) (hereafter SMT),
using a parametrization of the ellipsoidal collapse, showed that the threshold is mass dependent, and in particular that of the set of
objects that collapse at the same time, the less massive ones must initially have been denser than the more massive, since the less massive ones would have had to hold themselves together against stronger tidal forces.
In the second hand, the Gaussian nature of the fluctuation field means that a good approximation to the number density
of bound objects that have mass $m$ at time $z$ is given by considering the barrier crossing statistics of many independent and uncorrelated random walks, where the barrier shape $B(m,z)$, is connected to the collapse threshold.
Moreover, using the shape of the modified barrier in the excursion set approach, it is possible to obtain a good fit to the universal halo mass function. \footnote{Note that at present there is no good numerical test of analytic predictions for the low mass tail of the mass function.}

%NOTA
%SISTEMARE LA FIGURA 1 in modo che il mio fit sia meglio dell'altro

\begin{figure}
\label{Fig. 1}
\psfig{file=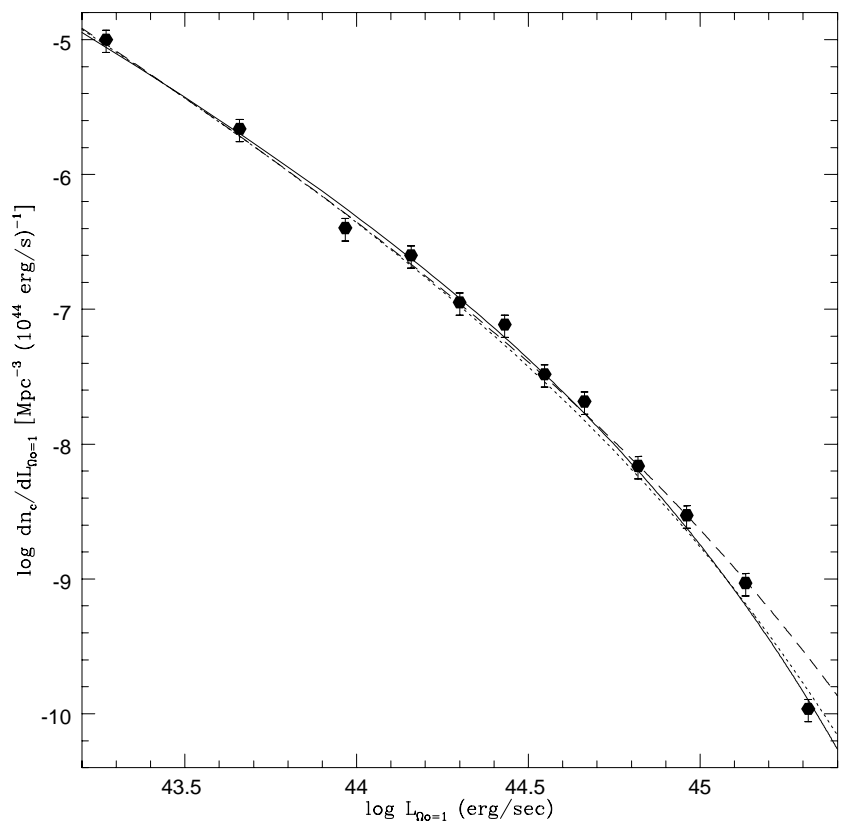,width=12cm}
%\resizebox{12cm}{!}{\includegraphics{fig1111.ps}}
%\parbox[b]{9cm}{
\caption{
The ROSAT BCS luminosity function. The solid line is the best fit of Eq. (17) of Reichart et al. (1999) to all 12 luminosity bins. The dotted line is the best fit of Eq. (\ref{eq:lfx}) to all 12 luminosity bins. The dashed line is the best fit of Eq. (\ref{eq:lfx}) to all but the highest-luminosity bin.}
%}
\end{figure}
As previously reported, the excursion set approach allows one to calculate good approximations to several
important quantities, such as the ``unconditional" and ``conditional" mass functions.
Sheth \& Tormen (2002) (hereafter ST) provided formulas to calculate these last quantities starting from the shape of the
barrier. They also showed that the
``unconditional"
%and ``conditional"
mass function, which is the one we need now, is in good agreement with results from numerical simulations.
Using the barrier shape obtained in Del Popolo \& Gambera (1998), obtained from the parametrization of the nonlinear collapse discussed in that paper, together with the results of ST in order to study the ``unconditional" and ``conditional" mass function.
\begin{figure}
\label{Fig. 1}
\psfig{file=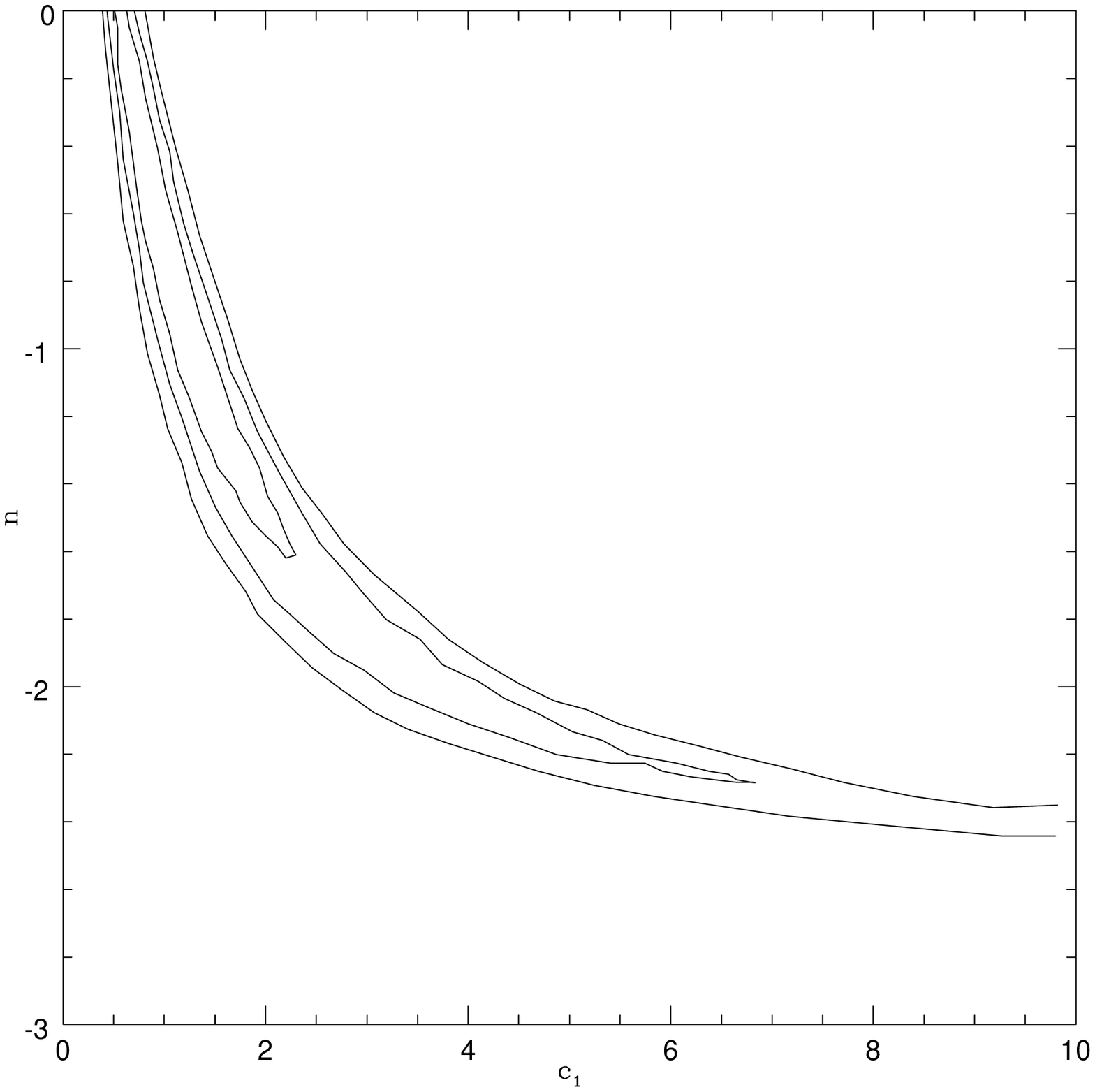,width=12cm}
%\resizebox{12cm}{!}{\includegraphics{ncc1.ps}}
%\parbox[b]{9cm}{
\caption{
The 1, 2 and $3 \sigma$ credible regions of the posterior probability distributions of the fit of
Eq. (\ref{eq:lfx}) to all but the highest-luminosity bin.}
%}
\end{figure}
Assuming that the barrier is proportional to the threshold for the collapse,
similarly to ST, the barrier can be expressed in the form:
\begin{equation}
B(M)=\delta _{\rm c}=\delta _{\rm co}\left[ 1+
\int_{r_{\rm i}}^{r_{\rm ta}}  \frac{r_{\rm ta} L^2 \cdot {\rm d}r}{G M^3 r^3}%
\right] \simeq \delta _{\rm co} \left[
1+\frac{\beta_1}{\nu^{\alpha_1}}
\right]
\label{eq:ma7}
\end{equation}
where $\delta _{\rm co}=1.68$ is the critical threshold for a spherical model,
$r_{\rm i}$ is the initial radius, $r_{\rm ta}$ is the turn-around radius,
$L$ the angular momentum, $\alpha_1=0.585$ and $\beta_1=0.46$.
%(for $\nu>0.1)$.
\footnote{The angular momentum appearing in Eq. ~(\ref{eq:ma7}) is the total angular momentum acquired by the proto-structure during evolution. In order to calculate $L$, I'll use the same model
as described in Del Popolo \& Gambera (1998, 1999) (more hints on
the model and some of the model limits can be found in
Del Popolo, Ercan \& Gambera (2001)).}
%Putting Eq. (\ref{eq:ma7}) into Eqs. ~(\ref{eq:distrib})-(\ref{eq:expans}) and truncating the expansion at $n=5$,
\begin{figure}
\label{Fig. 1} (a)
\resizebox{8cm}{!}{\includegraphics{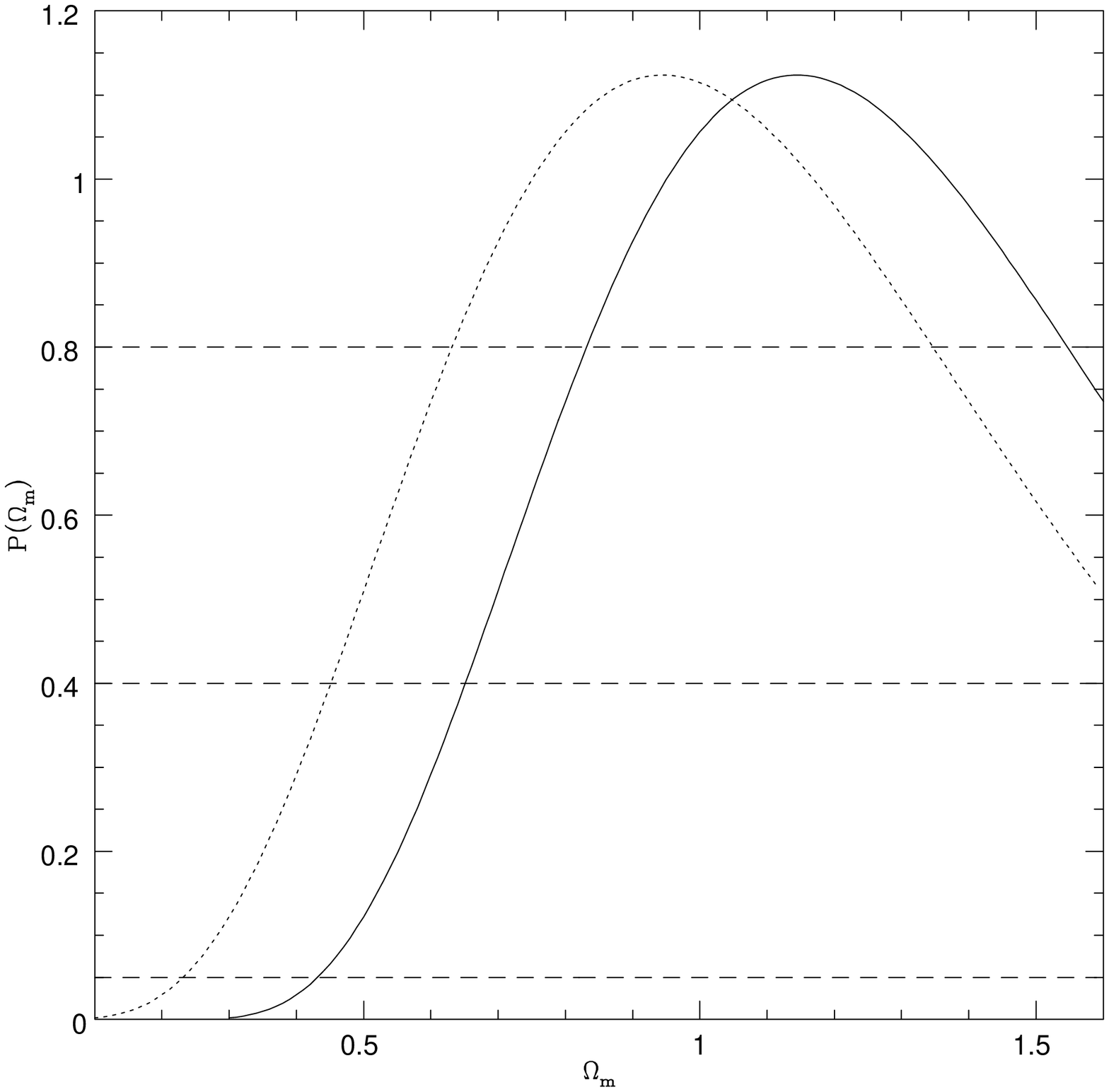}} (b)
\resizebox{8cm}{!}{\includegraphics{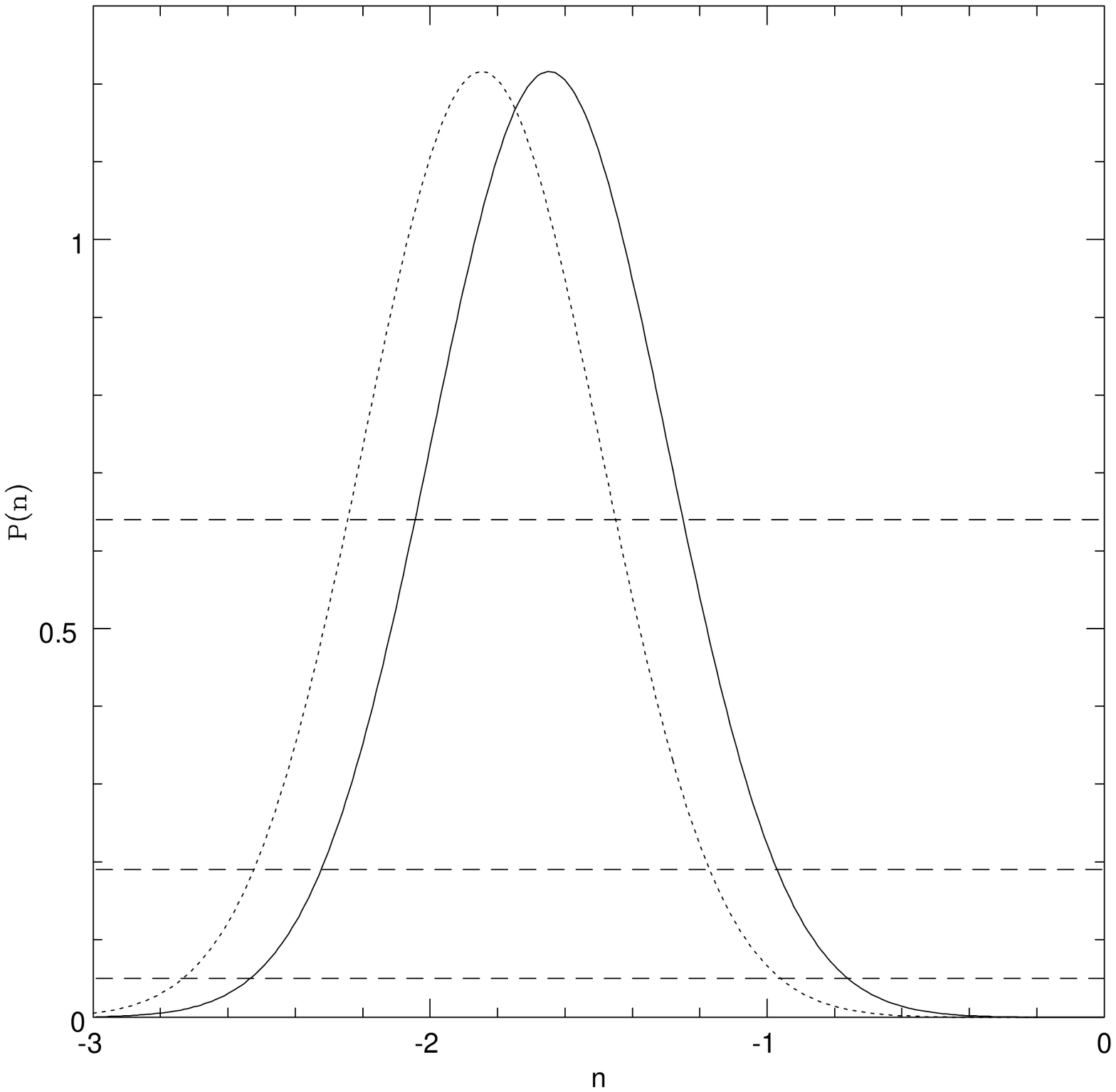}}
%\parbox[b]{9cm}{
\caption{
The marginalized posterior probability distributions $P(\Omega_{\rm m})$ (left panel) and $P(n)$ (right panel)
of the fit of Eq. (\ref{eq:cumul}) to the $0.14<z<0.6$ revised EMSS clusters and the ROSAT BCS luminosity function.
The solid line is the result of the calculation of this paper, the dotted line that of Reichart et al. (1999), and the dashed lines mark the 1, 2, 3 $\sigma$ credible intervals.
}
%}
\end{figure}
As described in Del Popolo (2000b), the
%``unconditional"
mass function can be approximated by:
%
%%\begin{equation}
%%f(\nu )d\nu \simeq 1.21\left( 1+\frac{0.06}{\left( a\nu \right) ^{0.585}}\right) \sqrt{\frac{a\nu }{2\pi }}\exp{\{-a\nu %%\left[ 1+\frac{0.57}{\left( a\nu \right) ^{0.585}}\right] ^{2}/2\}}
%%\end{equation}
%%where $a=0.707$
%
\begin{equation}
n(m,z)\simeq 1.21 \frac{\overline{\rho}}{m^{2}}\frac{d\log (\nu )}{d\log m}
\left( 1+\frac{0.06}{\left( a\nu \right) ^{0.585}}\right) \sqrt{\frac{a\nu }{2\pi }}\exp{\{-a\nu \left[ 1+\frac{0.57}{\left( a\nu \right) ^{0.585}}\right] ^{2}/2\}}
\label{eq:nmm}
\end{equation}
where $a=0.707$.
%Another important point in the calculation of the XLF is the $L_{\rm X}-M$
Eq. (\ref{eq:nmm}) can be converted from a mass function to a luminosity function. Following Mathiesen \& Evrard (1998) and Reichart et al. (1999), I assume that the bolometric luminosities are given by:
\begin{equation}
L_{\rm Bol} \propto M^p (1+z)^s
\end{equation}
and that:
\begin{equation}
L_{\rm X} \propto f_{\rm X} T^{-\beta} L_{\rm Bol}
\label{eq:lx}
\end{equation}
where $f_{\rm X}=0.989 \pm 0.014$ and $\beta=0.407 \pm 0.008$ for the representative temperature range of {\it EMSS} ($3 \leq T \leq 10 {\rm keV}$) and $f_{\rm X}=1.033 \pm 0.012$ and $\beta=0.472 \pm 0.008$ for the representative temperature range of {\it BCS} ($1.5 \leq T \leq 12 {\rm keV}$). In order to remove the temperature dependence introduced by
Eq. (\ref{eq:lx}), we need an M-T relation.
The M-T relation that I'll use, obtained in Del Popolo (2000a), is based on the merging-halo formalism of Lacey \& Cole (1993), accounting for the fact that
massive clusters accrete matter quasi-continuously, and is an improvement of a model proposed by
Voit (2000) (hereafter V2000), again to take account of angular momentum acquisition by protostructures.

%The evolution of the temperature distribution is very sensitive to the adopted model of structure
%formation, to the M-T relation, and its evolution at moderate redshift is considered a crucial test for cosmological models %(Kitayama \& Suto 1996).
%
%Recent observational improvement on the XTF was made
%by Markevitch (1998), Henry (2000),
%Blanchard et al. (2000), and Pierpaoli et al. (2001),
%using more accurate temperature-measurement results
%for each cluster with {\it ASCA} data (Tanaka et al. 1994).

%
%%As previously reported, in order to estimate the XTF of real systems  one needs
%%to know the temperature-mass (M-T) relation in
%%order to transform the mass distribution into the temperature distribution.

%%Theoretical uncertainty arises in this transformation because the exact
%%relation between the mass
%%appearing in the PS expression (or similar) and the temperature of the intra-cluster
%%gas is unknown.
%
Under the standard assumption of the
Intra-Cluster (IC) gas in hydrostatic equilibrium with the potential well
of a spherically symmetric, virialized cluster, the IC gas temperature-mass
relation is easily obtained by applying the virial
theorem
and for a flat matter-dominated Universe
it is given by (Evrard 1990, Evrard et al. 1996, Evrard 1997):
\begin{equation}
T=(6.4h^{2/3}keV)\left( \frac M{10^{15}M_{\odot }}\right) ^{2/3}(1+z)
\label{eq:ma2}
\end{equation}
The assumptions of perfect hydrostatic equilibrium and virialization
are in reality not completely satisfied in the case of clusters. Clusters profile may
depart from isothermality, with slight temperature gradients
throughout the cluster (Komatsu \& Seljak 2001). The X-ray weighted temperature can be slightly
different from the mean mass weighted virial temperature.
A noteworthy drawback of previous analyzes
has been stressed by Voit \& Donahue (1998) (hereafter V98)
and Voit (2000) (hereafter V2000). Using the merging-halo formalism of Lacey \& Cole (1993),
which accounts for the fact that massive clusters accrete matter quasi-continuously, they showed that
the M-T relation evolves, with time, more modestly than what expected in previous models predicting $T \propto (1+z)$,
and this evolution is even more modest in open universes.
Moreover, recent studies have shown that the self-similarity in the M-T relation seems to break at
some keV (Nevalainen, Markevitch \& Forman (hereafter NMF); Xu, Jin \& Wu 2001). By means of ASCA data,
using a small sample of 9 clusters (6 at 4 keV and 3 at $\sim 1$ keV), NMF has shown that $M_{\rm tot} \propto T_{\rm X}^{1.79 \pm 0.14}$ for the whole sample, and
$M_{\rm tot} \propto T_{\rm X}^{3/2}$ excluding the low-temperature clusters.
Xu, Jin \& Wu (2001) has found $M_{\rm tot} \propto T_{\rm X}^{1.60 \pm 0.04}$ (using the $\beta$ model), and $M_{\rm tot} \propto T_{\rm X}^{1.81 \pm 0.14}$ by means of the Navarro, Frenk \& White (1995) profile.
Finoguenov, Reiprich \& B\"oeringer (2001), have investigated the M-T relation in the low-mass end finding that $M \propto T^{\sim 2}$, and $M \propto T^{\sim 3/2}$ at the high mass end. This behavior has been attributed to the effect of the formation redshift (Finoguenov, Reiprich \& B\"oeringer 2001) (but see Mathiesen 2001 for a different
point of view), or to cooling processes (Muanwong et al. 2001) and heating (Bialek, Evrard \& Mohr 2000).
Afshordi \& Cen (2001) (hereafter AC) have shown that non-sphericity introduces an asymmetric, mass dependent,
scatter for the M-T relation altering its slope at the low mass end ($T \sim 3$ keV).\\
Clearly this has effects on the final shape of the temperature function.
In the following, I'll use a modified version of the M-T relation obtained
improving V98, V2000, to take account of tidal interaction between clusters. This M-T relation is given by:
\begin{equation}
kT \simeq 8 keV \left(\frac{M^{\frac 23}}{10^{15}h^{-1} M_{\odot}}\right)
\frac{
\left[
\frac{1}{m_1}+\left( \frac{t_\Omega }t\right) ^{\frac 23}
+\frac{K(m_1,x)}{M^{8/3}}
\right]
}
{
\left[
\frac{1}{m_1}+\left( \frac{t_\Omega }{t_{0}}\right) ^{\frac 23}
 +\frac{K_0(m_1,x)}{M_{0}^{8/3}}
\right]
}
\label{eq:kTT1}
\end{equation}
(see Del Popolo 2000a for a derivation),
where $M_{0} \simeq 5 \times 10^{14} h^{-1} M_{\odot}$,
$t_\Omega =\frac{\pi \Omega_{\rm m}}{H_o\left( 1-\Omega_{\rm m}-\Omega _\Lambda \right) ^{\frac 32}}$,
$m_1=5/(n+3)$ (being $n$ the spectral index), and:
\begin{eqnarray}
K(m_1,x)&=&F x \left (m_1-1\right ) {\it LerchPhi}
(x,1,3m_1/5+1)-
\nonumber \\
& &
F \left (m_1-1\right ){\it LerchPhi}(x,1,3m_1/5)
\end{eqnarray}
where $F$ is defined in the Del Popolo (2000a, b) and the
{\it LerchPhi} function is defined as follows:
\begin{equation}
LerchPhi(z,a,v)=\sum_{n=0}^{\infty} \frac{z^n}{(v+n)^a}
\end{equation}
where $x=1+(\frac{t_{\Omega}}{t})^{2/3}$ which is connected
to mass by $M=M_{\rm 0} x^{-3 m/5}$ (V2000),
and where $K_0(m_1,x)$ indicates that $K(m_1,x)$ must be calculated assuming $t=t_0$.

Eq. (\ref{eq:kTT1}) accounts for the fact that massive clusters accrete matter quasi-continuously, and takes account of tidal interaction between clusters.
The obtained M-T relation is no longer self-similar, a break in the low mass
end ($T \sim 3-4 {\rm keV}$) of the M-T relation is present.
The behavior of the M-T relation is as usual, $M \propto T^{3/2}$,
at the high mass end, and $M \propto T^{\gamma}$, with a value of
$\gamma>3/2$ in dependence of the chosen cosmology.
Larger values of $\gamma$ are related to open cosmologies, while
$\Lambda$CDM cosmologies give results of the slope intermediate
between the flat case and the open case.

%
%%It is given by:
%%\begin{equation}
%%kT \simeq 8 keV \left(\frac{M^{\frac 23}}{10^{15}h^{-1} M_{\odot}}\right)
%%\frac{
%%\left[
%%\frac 1m_1+\left( \frac{t_\Omega }t\right) ^{\frac 23}
%%+\frac{K(m_1,x)}{M^{8/3}}
%%\right]
%%}
%%{
%%\left[
%%\frac 1m_1+\left( \frac{t_\Omega }{t_{0}}\right) ^{\frac 23}
%%+\frac{K_0(m_1,x)}{M_0^{8/3}}
%%\right]
%%}
%%\label{eq:kT1}
%%\end{equation}
%%where we have defined $m_1=5/(n+3)$ (where $n$ is the usual power-law perturbation index), $t_\Omega =\frac{\pi \Omega %%_0}{H_o\left( 1-\Omega _0-\Omega _\Lambda \right) ^{\frac 32}}$ and $x=1+(\frac{t_{\Omega}}{t})^{2/3}$ which is connected %%to mass by $M=M_{\rm 0} x^{-3 m/5}$ (V2000).
%%where $K_0(m_1,x)$ indicates that $K(m_1,x)$ (see Del Popolo (2000b) for a definition of this quantity)
%%must be calculated assuming $t=t_0$.
By means of the previous M-T relation,  Eq. (\ref{eq:lx}) reads:
\begin{equation}
L_{\rm X} \propto
%c_1
f_{\rm X} M^{p-2\beta/3}
\left[\frac{
\left[
\frac 1m_1+\left( \frac{t_\Omega }t\right) ^{\frac 23}
+\frac{K(m_1,x)}{M^{8/3}}
\right]
}
{
\left[
\frac 1m_1+\left( \frac{t_\Omega }{t_{0}}\right) ^{\frac 23}
 +\frac{K_0(m_1,x)}{M_0^{8/3}}
\right]
}
\right]
^{-\beta}  (1+z)^s
\label{eq:lxx}
\end{equation}
where $p=1.77$ and $s=3.14-0.65 \Omega_{\rm m}$, for {\it EMSS}, and $p=1.86$ for {\it BCS} (Reichart, Castander \& Nichol 1998; Reichart et al. 1999). \footnote{Since the {\it BCS} is a local ($z_{\rm eff} \simeq 0.1$) catalog, the value of $s$ is unimportant.}
For $t_\Omega>>t$, Eq. (\ref{eq:lxx}) reduces to:
\begin{equation}
L_{\rm X} \propto
%c_1
f_{\rm X} M^{p-2\beta/3} (1+z)^{s-\beta}
\label{eq:lxxx}
\end{equation}
(see Reichart et al. 1999, Eq. 11),
since the M-T relation reduces to the standard M-T relation ($T \propto M^{2/3} (1+z)$).
%{\it late-formation} formula (see V2000)????
Taking account only the change in the mass function, substituting Eq. (\ref{eq:lxxx}) into
Eq. (\ref{eq:nmm}), defining $L_{\rm X}=x(z)L_1$\footnote{The {\it EMSS} and {\it BCS} provide luminosities in their respective X-ray bands that have been computed for $H_0=50 {\rm km Mpc^{-1} s^{-1}}$ and $\Omega_{\rm m}=1$. These luminosities are denoted by $L_1$} (see Reichart et al. 1999 for a definition of $x(z)$), and assuming a scale-free mass density fluctuation power spectrum of power law index $n$, so that the present variance is:
\begin{equation}
\sigma_0(M)=\sigma_8 (\frac{M}{M_8})^{\frac{-(3+n)}{6}}
%c_2M^{\frac{-(3+n)}{6}}
\label{eq:sigm}
\end{equation}
where $M_8=\frac{H_0^2 \Omega_{\rm m}}{2 G} (\rm 8 h^{-1} Mpc)^3=6.0 \times 10^{14} \Omega_{\rm m} h^{-1} M_{\odot}$ (Voit 2000) ($\simeq 10^{15} M_{\odot}$, in a flat universe with $h=0.5$).
\footnote{$\sigma_8$ is the amplitude of the mass density fluctuation power spectrum over spheres of radius $8 h^{-1} {\rm Mpc}$, and $M_8$ is the mean mass within these spheres}
I obtain the following luminosity function:

\begin{equation}
\frac{dn_{\rm c}}{d L_1}=
%-A\frac{\sqrt{2}}{2c_{2}}\sqrt{\frac{a}{\pi }}\delta \rho \frac{3+n}{2b-3p}
c_2 F(z,L_1) \exp\left[-G(z,L_1)\right]
\label{eq:lumin}
\end{equation}

\begin{maplelatex}
\[
{ F(z,L_1)}= \,\left[{ f_1(z)}\,{ L_1}^{(1/2\,\frac { - 3 + n}{3\,p
 - 2\,\beta} - 1)} + { f_2(z)}\,{ L_1}^{( - 1/2\,\frac {3 - n + 6\,
\alpha  + 2\,\alpha \,n}{3\,p - 2\,\beta} - 1)}\right]
\]
\end{maplelatex}

\begin{maplelatex}
\[
{ f_1(z)}=
{{c_1}}^{1/2\,{\frac {-3+n}{-3\,p+2\,b}}}
\delta_{\rm c0}(z) \,(1 + z)^{\frac {(s - \beta)\,(3 - n)}{6\,p - 4
\,\beta}}
\,{ f_{\rm X}}^{ - 1/2\,\frac { - 3 + n}{3\,p - 2\,\beta}} x(z)^{1/2\,\frac { - 3 + n}{3\,p
 - 2\,\beta}}
\]
\end{maplelatex}

\begin{maplelatex}
\[
{ f_2(z)}=
{{c_1}}^{-1/2\,{\frac {3-n+6\,\alpha+2\,\alpha\,n}{-3\,p+2\,b}}}
\delta_{\rm c0}(z)^{(1 - 2\,\alpha )}\,(1 + z)^{\left[ - 1/2\,
\frac {( - s + \beta)\,(3 - n + 6\,\alpha  + 2\,\alpha \,n)}{3\,p - 2
\,\beta}\right]}\,{ f_{\rm X}}^{1/2\,\frac {3 - n + 6\,\alpha  + 2\,\alpha \,n
}{3\,p - 2\,\beta}}\,a^{ - \alpha } 
\]
\end{maplelatex}
\begin{maplelatex}
\[
a_1 x(z)^{- 1/2\,\frac {3 - n + 6\,
\alpha  + 2\,\alpha \,n}{3\,p - 2\,\beta}}
%c_2^{2\alpha}
\]
\end{maplelatex}

\begin{maplelatex}
\[
{ G(z,L_1)}=  \left[{ g_1(z)}\,{ L_1}^{\frac {3 + n}{3\,p - 2\,\beta}} +
{ g_2(z)}\,{ L_1}^{ - \frac {(\alpha  - 1)\,(3 + n)}{3\,p - 2
\,\beta}} + { g_3(z)}\,{ L1}^{ - \frac {( - 1 + 2\,\alpha )\,(3
 + n)}{3\,p - 2\,\beta}}\right]
\]
\end{maplelatex}

\begin{maplelatex}
\[
{ g_1(z)}=\frac{1}{2}
%{\displaystyle
{{c_1}}^{{\frac {3+n}{-3\,p+2\,b}}}
\,{ f_{\rm X}}^{ - \frac {
3 + n}{3\,p - 2\,\beta}}\,(1 + z)^{ - \frac {(s - \beta)\,(3 + n)}{3\,p
 - 2\,\beta}}\,a\,\delta_{\rm c0}(z) ^{2} x(z)^{\frac {3 + n}{3\,p - 2\,\beta}}
\]
\end{maplelatex}

\begin{maplelatex}
\[
{ g_2(z)}=
{{c_1}}^{-{\frac {-3-n+3\,\alpha+\alpha\,n}{-3\,p+2\,b}}}
a^{(1 - \alpha )}\,\delta_{\rm c0}(z) ^{(2 - 2\,\alpha )}\,
{ f_{\rm X}}^{\frac {(\alpha  - 1)\,(3 + n)}{3\,p - 2\,\beta}}\,(1 + z)
^{\frac {(s - \beta)\,(3 + n)\,(\alpha  - 1)}{3\,p - 2\,\beta}}\,{
a_2} x(z)^{- \frac {(\alpha  - 1)\,(3 + n)}{3\,p - 2
\,\beta}}
%c_2^{2 \alpha}
\]
\end{maplelatex}

\begin{maplelatex}
\[
{ g_3(z)}=\frac{1}{2}
%{\displaystyle
{{c_1}}^{-{\frac { \left( 3+n \right)  \left( -1+2\,\alpha \right)
}{-3\,p+2\,b}}}
\,a^{(1 - 2\,\alpha
)}\,\delta_{\rm c0}(z) ^{(2 - 4\,\alpha )}\,{ f_{\rm X}}^{\frac {( - 1 + 2\,
\alpha )\,(3 + n)}{3\,p - 2\,\beta}}\,(1 + z)^{\frac {(s - \beta)\,(3
 + n)\,( - 1 + 2\,\alpha )}{3\,p - 2\,\beta}}\,{ a_2}^{2} x(z)^{- \frac {( - 1 + 2\,\alpha )\,(3
 + n)}{3\,p - 2\,\beta}}
%c_2^{4 \alpha}
\]
\end{maplelatex}
where $a_1=0.06$ and $a_2=0.57$ and the constants $c_1$ and $c_2$ corresponds to the constants $a$ and $c$ in Reichart et al. (1999), respectively. As in Reichart et al. (1999), $c_1$ and $c_2$ depend upon $\sigma_8$ and the factor of proportionality of Eq. (\ref{eq:lxx}).

Taking also account of the changes in the M-T relation, and using Eq. (\ref{eq:lxx}), the XLF function can be written as:

\begin{equation}
\frac{dn_{\rm c} (m,z)}{dL_{\rm X}}=\frac{A(m,z)}{B(m,z)}
\label{eq:lfx}
\end{equation}

\begin{eqnarray*}
%\lefteqn{{A}=  -
%{\displaystyle \frac {1}{6}} A\,\sqrt{
%2}\,\sqrt{a}\,\delta \,\rho \,(3 + n)} \\
% & &
A(m,z)&=&c_2  e^{( - 1/2\,m^{{m_{3}}}\,a\,\delta ^{2} - a^{(1 - \alpha )}
\,\delta ^{(2 - 2\,\alpha )}\,m^{({m_{3}} + 2\,{m_{4}}\,\alpha )}
\,{a_2} - 1/2\,a^{(1 - 2\,\alpha )}\,\delta ^{(2 - 4\,\alpha )
}\,m^{({m_{3}} + 4\,{m_{4}}\,\alpha )}\,{a_2}^{2})}\,(1 + z)^{
( - s)}((m^{r} \\
 & & \mbox{} + m^{r}\,{t_{1}}\,{m_{1}} + m^{(r - q + {a_3})}\,
F \cdot\,{LPh}\,{m_{1}}^{2} - m^{(r - q)}\,F \cdot\,{LPh}\,{m_{1}}^{2
} - m^{(r - q + {a_3})}\,F \cdot\,{LPh}\,{m_{1}} \\
 & & \mbox{} + m^{(r - q)}\,F \cdot\,{LPh}\,{m_{1}}) \left/
{\vrule height0.47em width0em depth0.47em} \right. \!  \! {m_{1}}
)^{\beta}( - m^{( - p + 1 + {m_{2}} + q + {a_3})} + m^{( - p + 1
 + {m_{2}} + q)} - m^{( - p + 1 + {m_{2}} + q + {a_3})}\,{t_{1
}}\,{m_{1}} \\
 & & \mbox{} + m^{( - p + 1 + {m_{2}} + q)}\,{t_{1}}\,{m_{1}} - m
^{( - p + 1 + {m_{2}} + 2\,{a_3})}\,F \cdot\,{LPh}\,{m_{1}}^{2}
 + 2\,m^{( - p + 1 + {m_{2}} + {a_3})}\,F \cdot\,{LPh}\,{m_{1}}
^{2} \\
 & & \mbox{} - m^{( - p + 1 + {m_{2}})}\,F \cdot\,{LPh}\,{m_{1}}^{2
} + m^{( - p + 1 + {m_{2}} + 2\,{a_3})}\,F \cdot\,{LPh}\,{m_{1}}
 - 2\,m^{( - p + 1 + {m_{2}} + {a_3})}\,F \cdot\,{LPh}\,{m_{1}}
 \\
 & & \mbox{} + m^{( - p + 1 + {m_{2}})}\,F \cdot\,{LPh}\,{m_{1}} -
m^{( - p + 1 + {m_{2}} + 2\,{m_{4}}\,\alpha  + q + {a_3})}\,{a
_{1}}\,a^{( - \alpha )}\,\delta ^{( - 2\,\alpha )} \\
 & & \mbox{} + m^{( - p + 1 + {m_{2}} + 2\,{m_{4}}\,\alpha  + q)}
\,{a_{1}}\,a^{( - \alpha )}\,\delta ^{( - 2\,\alpha )} - m^{( - p
 + 1 + {m_{2}} + 2\,{m_{4}}\,\alpha  + q + {a_3})}\,{a_{1}}\,a
^{( - \alpha )}\,\delta ^{( - 2\,\alpha )}\,{t_{1}}\,{m_{1}} \\
 & & \mbox{} + m^{( - p + 1 + {m_{2}} + 2\,{m_{4}}\,\alpha  + q)}
\,{a_{1}}\,a^{( - \alpha )}\,\delta ^{( - 2\,\alpha )}\,{t_{1}}\,
{m_{1}} \\
 & & \mbox{} - m^{( - p + 1 + {m_{2}} + 2\,{m_{4}}\,\alpha  + 2\,
{a_3})}\,{a_{1}}\,a^{( - \alpha )}\,\delta ^{( - 2\,\alpha )}
\,F \cdot\,{LPh}\,{m_{1}}^{2} \\
 & & \mbox{} + 2\,m^{( - p + 1 + {m_{2}} + 2\,{m_{4}}\,\alpha  +
{a_3})}\,{a_{1}}\,a^{( - \alpha )}\,\delta ^{( - 2\,\alpha )}
\,F \cdot\,{LPh}\,{m_{1}}^{2} \\
 & & \mbox{} - m^{( - p + 1 + {m_{2}} + 2\,{m_{4}}\,\alpha )}\,{a
_{1}}\,a^{( - \alpha )}\,\delta ^{( - 2\,\alpha )}\,F \cdot\,{LPh}
\,{m_{1}}^{2} \\
 & & \mbox{} + m^{( - p + 1 + {m_{2}} + 2\,{m_{4}}\,\alpha  + 2\,
{a_3})}\,{a_{1}}\,a^{( - \alpha )}\,\delta ^{( - 2\,\alpha )}
\,F \cdot\,{LPh}\,{m_{1}} \\
 & & \mbox{} - 2\,m^{( - p + 1 + {m_{2}} + 2\,{m_{4}}\,\alpha  +
{a_3})}\,{a_{1}}\,a^{( - \alpha )}\,\delta ^{( - 2\,\alpha )}
\,F \cdot\,{LPh}\,{m_{1}} \\
 & & \mbox{} + m^{( - p + 1 + {m_{2}} + 2\,{m_{4}}\,\alpha )}\,{a
_{1}}\,a^{( - \alpha )}\,\delta ^{( - 2\,\alpha )}\,F \cdot\,{LPh}
\,{m_{1}})
\end{eqnarray*}

\begin{eqnarray*}
B(m,z)&=& {m}^{3/2}
%\sqrt {\pi }
{ c_1 f_{\rm X}}\, ( \beta F{ a_3}\,{{ m_1}}^{2}-\beta F{
 a_3}\,{ m_1}-\beta {m}^{q}r+{m}^{q}p-pF{ LPh}  {{ m_1}}^{2}+pF{ LPh}  { m_1}+
\nonumber \\
& &
{m}^{q}p{ t_1}\,{ m_1}-{
m}^{q+{ a_3}}p-\beta F{ a_3}\,{{ m_1}}^{2}{ m_{1l}}\,{ LPh}
-\beta rF{ LPh} ) { m_1}+\beta r{{ m_1}}^{2}F{LPh}+
\beta F{ LPh}
  q{ m_1}+
\nonumber \\
& &
\beta F{ a_3}\,{
m_{1l}}\,{ LPh}  {m_1
}+
\beta {m}^{{ a_3}}F{ a_3}\,{ m_1}+\beta {m}^{{ a_3}}F{ LPh}
  { a_3}\,{ m_1}-\beta {m}^{q}
r{ t_1}\,{ m_1}-2\,{m}^{{ a_3}}pF{ LPh} { m_1}+
\nonumber \\
& &
2\,{m}^{{ a_3}}pF{ LPh
}  {{ m_1}}^{2}+
2\,\beta {m}^{{
 a_3}}F{ LPh}  {{
 m_1}}^{2}q-
\beta {m}^{{ a_3}}F{ LPh}  {{ m_1}}^{2}{ a_3}-2\,\beta {m}^{{ a_3}}F{
LPh}  q{ m_1}+
\nonumber \\
& &
2\,\beta {m}^
{{ a_3}}rF{ LPh}  {
 m_1}+
2\,\beta {m}^{{ a_3}}F{ a_3}\,{{ m_1}}^{2}{ m_{1l}}\,{
LPh}
-2\,\beta {m}^{{ a_3}
}F{ a_3}\,{ m_{1l}}\,{ LPh}  { m_1}-\beta {m}^{{ a_3}}F{ a_3}\,{{ m_1}}^{2}-
\nonumber \\
& &
\beta F{
LPh}  {{ m_1}}^{2}q-
2
\,\beta {m}^{{ a_3}}r{{ m_1}}^{2}F{ LPh}  +\beta {m}^{q+{ a_3}}r+{m}^{2\,{ a_3}}pF{
LPh}  { m_1}+
\beta {m}^{2\,
{ a_3}}F{ a_3}\,{ m_{1l}}\,{ LPh}  { m_1}+
\nonumber \\
& &
\beta {m}^{2\,{ a_3}}F{ LPh}
{{ m_1}}^{2}{ a_3}-{m}^{q+{ a_3
}}p{ t_1}\,{ m_1}+
\beta {m}^{2\,{ a_3}}F{ LPh}  q{ m_1}
+
\beta {m}^{2\,{ a_3}}r{{ m_1}}^{
2}F{ LPh}-
\nonumber \\
& &
\beta {m}^{2\,
{ a_3}}rF{ LPh}  {
 m_1}
+
\beta {m}^{q+{ a_3}}r{ t_1}\,{ m_1}-
\beta {m}^{2\,{ a_3}}F{
LPh}  {{ m_1}}^{2}q-
{m
}^{2\,{ a_3}}pF{ LPh}  {{ m_1}}^{2}-
\nonumber \\
& &
\beta {m}^{2\,{ a_3}}F{ LPh}  { a_3}\,{ m_1}-
\beta {m}^{2\,{ a_3}}F{
 a_3}\,{{ m_1}}^{2}{ m_{1l}}\,{ LPh})
\end{eqnarray*}
%NOTA
%IMPORTANTE: RICORDARE CHE $\sigma_8$ NON c'e' nella formula precedente e anche nella precedente-precedente

where, in order to simplify the previous expression, I have defined:
$m=\frac{M}{M_8}$,
%\footnote{Note that now $m$ in no longer the parameter of m=5/(n+3), previously seen},
$m_1=5/(n+3)$,
$m_2=n/6$, $m_3=1+n/3$, $m_4=-1/2-n/6$, $a$, $a_1$, and $a_2$ are the same of Eq. (\ref{eq:lumin}
), $a_3=-5/(3m1)$, $m_{1l}=3m1/5+1$, $t_1=(t_{\Omega}/t)^{2/3}$, $r=2/3$, $q=8/3$, $\delta_{\rm c0}(z)=\delta$,
and $LerchPhi(m^{a3},1,m_{1l})=LPh$.

%
%%SISTEMARE LE COSTANTI, SONO ERRATE NELLA PRIMA E MANCANO NELLA SECONDA. NELLA SECONDA MANCA ANCHE x(z)
%

Then the luminosity function can be written in implicit form as:
~\\
~\\

%
%%\begin{eqnarray*}
%%Lx&=&{ f_{\rm X}}\,\left ({\frac {{m}^{r}\left (1+{ t_1}\,{ m1}+{m}^{-q+{
%%a_3}}FLPh{{ m1}}^{2}-{m}^
%%{-q}FLPh{{ m1}}^{2}-{m}^{-q+
%%{ a_3}}FLPh{ m1}+{m}^{-q}F
%%LPh{ m1}\right )}{{ m1}}}
%%\right )^{-b} \\
%% & & \times {m}^{p}\left (1+z\right )^{s}
%%\end{eqnarray*}
%
%%%%%\[
%%%%%\left\{
%%%%%\begin{array}{lc}

\begin{equation}
\frac{dn_{\rm c} (m,z)}{dL_{\rm X}}=\frac{A(m,z)}{B(m,z)}
\label{eq:prim}
\end{equation}

\begin{eqnarray*}
L_{\rm X}=x(z) L_1&=& c_1 { f_{\rm X}}\,({m}^{r} (1+{ t_1}\,{ m_1}+{m}^{-q+{
a_3}}FLPh{{ m_1}}^{2}-{m}^
{-q}FLPh{{ m_1}}^{2}-{m}^{-q+
{ a_3}}FLPh{ m_1}+ 
\\
& &
{m}^{-q}F
LPh{ m_1}) 
)^{-\beta} 
%\\
% & & 
%\times 
{m}^{p} m_1^{\beta}\left (1+z\right )^{s}
\end{eqnarray*}
%%%%%%\end{array}
%%%%\right.
%%%%\]

The total number of X-ray clusters observed between luminosity and redshift limits $L_{\rm l}<L1<L_{\rm u}$ and
$z_{\rm l}<z<z_{\rm u}$, i.e. the cumulative luminosity function is given by:
\begin{equation}
N(L_{\rm l}, L_{\rm u}, z_{\rm l}, z_{\rm u})= \int_{L_{\rm l}}^{L_{\rm u}}\int_{z_{\rm l}}^{z_{\rm u}}
A(L_1,z) \frac{dn_{\rm c} (L_1,z)}{dL_1} d L_1 dV(z)
\label{eq:cumul}
\end{equation}
where $A(L_1,z)$ is the area of the sky that an X-ray survey samples at redshift $z$ as a function of luminosity $L_1$.
In the case of the {\it EMSS}, this quantity is given by (Avni \& Bahcall 1980; Henry et al. 1992, Nichol et al. 1997)
\begin{equation}
A(L_1,z)=A(F_{\rm lim}=F(L_1,z))
\end{equation}
where $A(F_{\rm lim})$ is the area that
the {\it EMSS} surveyed below sensitivity limit $F_{\rm lim}$ (see Henry et al. 1992):
\begin{equation}
F(L_1,z)=\frac{f_{\rm F}(d_{\rm A(z)})}{k(z)} \frac{h^2_{50}L_1}{4 \pi d^2_{\rm L}(z)}
\end{equation}
being $k(z)$ the -correction from the observer frame to the source frame for a $T=6 {\rm keV}$ X-ray cluster. $A(L_1,z)$ can be calculated similarly to Reichart et al. (1999) (see their Fig. 1).
Finally
\begin{equation}
dV(z)=\frac{4c^{3}dz}{H_{0}^{3}\Omega _{m}^{4}(1+z)^{3}}\frac{(\Omega _{m}z+(\Omega _{m}-2)((\Omega _{m}z+1)^{\frac{1}{2}}-1))^{2}}{\left( 1+\Omega _{m}z\right) ^{\frac{1}{2}}}
\end{equation}
is the comoving volume element.
The previous model, has reported by Reichart et al. (1999) consists of the parameters: $H_0$, $f_{\rm X}$, $\beta$, $p$, $s$, $c_1$, $c_2$, $n$ and $\Omega$. Fixing the value of $H_0$, using the quoted values of $f_{\rm X}$, $\beta$, $p$, $s$, and noticing that the normalization parameter, $c_2$, drops out
of the Bayesian inference analysis of this paper (see Eq. (\ref{eq:prob})),
%(ma non e' vero),
we are left with the parameters $\Omega_{\rm m}$, $n$, $c_1$.
It is possible to obtain credible intervals for the values of the quoted parameters as described in Reichart et al. (1999).
As in the quoted paper, the parameter $n$ and $c_1$ are constrained with the local $z_{\rm eff} \simeq 0.1$ luminosity function of the BCS. Althought such a sample may not have enough redshift leverage to adequately probe $\Omega_{\rm m}$, its large size makes it an excellent sample to constraint the parameters $n$ and $c_1$. These constraints can be combined with
%$f(z)=f(z_{\rm eff})=f_(\rm eff)$ and $f(z)=f(z_{\rm eff})=f_(\rm eff)$
the {\rm EMSS} results to better constrain $\Omega_{\rm m}$. So letting $z=z_{\rm eff}$, $A(L_1,z_{\rm eff})=1$ (see
Reichart et al. 1999 for a discussion), using the binned BCS luminosity function of Fig. 1 of Ebeling et al. (1997),
I find the parameters by using a likelihood function given by $e^{-\chi^2/2}$, with $-3<n<0$ and
%$0<g_{\rm eff}<10$.
$0<c_1<10$.
Fitting Eq. (\ref{eq:prim}) to all data, I obtain $n=-0.65^{+0.3}_{-0.28}$ and $c_1=0.77^{+0.25}_{-0.11}$.
When fitting Eq. (\ref{eq:prim}) to all but the highest luminosity bin, I find that
$n=-1.53^{+0.8}_{-0.12}$
%$n=-1.61^{+0.8}_{-0.12}$
and $c_1=1.05^{+0.65}_{-0.55}$.
This double fitting is due to safeguard against a bias on $n$ that can be introduced by the assumption $z=z_{\rm eff}$ (see Reichart et al. 1999 for a deeper description).

In Fig. 1, I plot
the ROSAT BCS luminosity function. The solid line is the best fit of Eq. (17) of Reichart et al. (1999) to all 12 luminosity bins. The dotted line is the best fit of Eq. (\ref{eq:lfx}) to all 12 luminosity bins. The dashed line is the best fit of Eq. (\ref{eq:lfx}) to all but the highest-luminosity bin.
%
%I plot the ROSAT BCS luminosity function.
%The solid line is the best fit of Eq. (17) of Reichart et al. (1999) to all 12 luminosity bins. The dotted line is the best %fit of Eq. (\ref{eq:lfx}) to all 12 luminosity bins. The dashed line is the best fit of Eq. (\ref{eq:lfx}) to all but the %highest-luminosity bin
%of Ebeling et al. (1997). The solid line is the best-fit Schechter function of Ebeling et al. (1997). The dotted line is %the best fit of Eq. (\ref{eq:prim}) to all bins, and the dashed line the same fit to all but the highest-luninosity bin.
%
In Fig. 2, I plot the 1, 2, $3 \sigma$ credible regions in the n-$c_1$ plane for the second fit (all but the highest luminosity bin).
%\footnote{The credible regions in the n-$c_2$ ??????????????????????????}

%SERVE LA FIGURA 2 DI REICHART PERCHE? E? USATA PER OTTENERE LA EMSS/BCS COMBINATA
As previously described, the data coming from {\rm EMSS} are used to obtain constraints on $\Omega_{\rm m}$, $n$, $c_1$, by means of Bayesian inference analysis.
In order to perform the quoted analysis, we need
the posterior probability distribution for $\Omega_{\rm m}$, $n$, $c_1$, namely $P(\Omega_{\rm m},n,c_1)$
which
is obtained by normalizing the product of the prior probability distribution and the likelihood function (Gregory \& Loredo 1992). The prior assumed is flat between $0<\Omega_{\rm m}<1.5$, $-3<n<0$ and $0<c_1<3$.
In order to constraint, $\Omega_{\rm m}$, $n$ and $c_1$, I use the likelihood function $L(\Omega_{\rm m},n,c_1)$ (see Cash 1979, Reichart et al. 1999):
\begin{equation}
L(\Omega_{\rm m},n,c_1)=\prod_{i\equiv 1}^{N_{tot}}P(L_{1},_{i},z_{i}|\Omega _{m},n,c_1)
\label{eq:likel}
\end{equation}
where $P(L_{1},_{i},z_{i}|\Omega _{m},n,c_1)$ is the probability that the $i$th X-ray cluster fits our model, given values of
$\Omega _{\rm m}$, $n$, $c_1$. For our model, this probability is given by (e.g., Cash 1979):
\begin{equation}
P(L_1,z|\Omega _{m},n,c_1)=
\frac{A(L_1,z) \frac{dn_{\rm c} (L_1,z)}{dL_1} \frac{dV(z)}{dz}}{N(L_{\rm l}, L_{\rm u}, z_{\rm l}, z,_{\rm u})}
\label{eq:prob}
\end{equation}
In Eq. (\ref{eq:likel}), $N_{\rm tot}$ is the total number of X-ray clusters in the same region of the $L_1-z$ plane as that over which $N(L_{\rm l}, L_{\rm u}, z_{\rm l}, z,_{\rm u})$ is defined. The value of $L_{\rm l}<L_1<L_{\rm u}=10^{45.5} {\rm erg/s}$. $L_{\rm l}$ is set by the limiting flux of the {\it EMSS}: $F(L_{\rm l},z)=1.33 \times 10^{-13} {\rm erg/cm^2 s}$ (Henry et al. 1992). The redshift is in the range: $0.14=z_{\rm l}<z<z_{\rm u}=0.6$.

%$z_u=0.9$?????

The posterior probability distribution for one of the parameters, e.g., $P(\Omega_{\rm m})$, is obtained by marginalizing the posterior probability distribution for all the three parameters, $P(\Omega_{\rm m},n,c_1)$, over the other parameters (Gregory \& Loredo 1992). The credible regions (1, 2, 3 $\sigma$) are determined by integrating the posterior probability distribution over the most probable region of its parameter space until 68,3\%, 95.4\% and 99.73\% (respectively) of this distribution has been integrated.
In Fig. 3, I plot
%the marginalized posterior probability distribution of $P(\Omega_{\rm m})$ and $P(n)$ of the fit of
%Eq. (\ref{eq:cumul}) to the $0.14<z<0.6$ revised EMSS clusters and the ROSAT BCS luminosity function.
the marginalized posterior probability distributions $P(\Omega_{\rm m})$ (left panel) and $P(n)$ (right panel)
of the fit of Eq. (\ref{eq:cumul}) to the $0.14<z<0.6$ revised EMSS clusters and the ROSAT BCS luminosity function.
The solid line is the result of the calculation of this paper, the dotted line that of Reichart et al. (1999).
%, and the dashed lines mark the 1, 2, 3 $\sigma$ credible intervals.

%the 1, 2, 3$\sigma$ credible regions of the combined {\it EMSS/BCS} posterior probability distributions, $P(\Omega_{\rm %m})$ and $P(n)$, calculated as described in Reichart et al. (1999).
\footnote{Namely, the likelihood function is that of the {\it EMSS}, Eq. (\ref{eq:likel}). Instead of assuming a flat priori distribution for all three parameters, I assume a flat priori between $0<\Omega_{\rm m}<1.5$, and use the posterior probability distribution of the BCS-$P_{BCS}(n,c_1)$ (see Fig. 2)-Eq. (\ref{eq:lumin}), and the effective redshift of the BCS, $z_{\rm eff} \simeq 0.1$, to determine the full prior probability distribution: $P_{BCS} (\Omega_{\rm m},n,c_1)$.}
In Fig. 3a-b, the dashed lines represents the 1, 2, and $3 \sigma$ credible intervals, the dotted line of Fig. 3a represents
$P(\Omega_{\rm m})$ versus $\Omega_{\rm m}$, while Fig. 3b represents $P(n)$ versus $n$.
The figure
%and it
shows that $\Omega_{\rm m}=1.15^{+0.40}_{-0.33}$ and
$n=-1.55^{+0.42}_{-0.41}$.
%$n=-1.63^{+0.42}_{-0.41}$.
The previous result shows that the change in the mass function and M-T relation gives rise to an increase of $\Omega_{\rm m}$ and $n$ of $\simeq 20\%$.
%, while the change in $n$ is $\simeq 20\%$.
%(AUMENTARE!).

The lesson from the previous calculation is that taking account of non-sphericity in collapse and the fact that
massive clusters accrete matter quasi-continuously gives rise to a noteworthy change in the prediction of cosmological parameters, as $\Omega_{\rm m}$. Here, I am principally interested in studying the effects of these changes on the values of cosmological parameters, and not in the peculiar value obtained. In this light,
%other terms,
the large value of $\Omega_{\rm m }$ obtained is not of particular importance or meaning, but is simply a consequence of the high value of $\Omega_{\rm m }$ obtained in Reichart et al. (1999) analysis.
%
%%\footnote{A possible explanation to the high value of $\Omega_{\rm m}$ obtained by Reichart et %%al. (1999) is that the EMSS missed many high-redshift, high-luminosity X-ray clusters (see %%Reichart et al. (1999)).}
%
%Commento........($\Omega_{\rm m}=1.15^{+0.40}_{-0.33}$ e' troppo grande....)
As an example, I have also estimated the value of $\Omega_{\rm m }$ following Borgani et al. (2001).
Analyzing the ROSAT Deep Cluster Survey (RDCS) and using the XLF to obtain constraints on cosmological parameters, Borgani et al. (2001), found that $\Omega_{\rm m}=0.35^{+0.13}_{-0.10}$. In their study, they used the ST mass function instead of the usual PS and Eke et al. (1998) M-T relation.
Using their method and data, but our mass function and M-T relation, one obtains larger values of $\Omega_{\rm m}$ ($\Omega_{\rm m} \simeq 0.4 \pm 0.1$) that differently from the previous analysis (Reichart et al. 1999) exclude an Einstein-de Sitter model.
Note that since the RDCS data are not avaliable to date, I used data taken from the XLF in Rosati, Borgani \& Norman, (2002).
In any case, even using all RDCS data small changes are expected, and the quoted result, namely that
the Einstein-de Sitter model is excluded, is not changed, but a larger value of $\Omega_{\rm m}$
is obtained.
 %Aggiungere i risultati di Borgani e quelli miei relativi usando il metodo di Borgani......

%CHIEDERE I DATI A BORGANI

\section{Constraints to $\Omega_{\rm m}$, $n$, and $\sigma_8$ from the XTF}

%The haloes mass functions of clusters of galaxies
%contains information on the structure formation history of the universe: they are a primary input
%for modeling galaxy formation.
As previously reported, the mass function (MF) is a critical ingredient in putting strong constraints on
cosmological parameters (e.g., $\Omega_{\rm m}$).
%and $\Lambda$).
Observationally the local mass function has been derived from
measuring masses of individual clusters from galaxy velocity dispersions or
other optical properties by Bahcall and Cen (1993), Biviano et al. (1993), and
Girardi et al. (1998). However, the estimated virial masses for individual clusters
depend rather strongly on model assumptions.
As argued by Evrard et al. (1996) on the basis of hydrodynamical
N-body simulations, cluster masses may be presently more accurately
determined from a temperature measurement
and a mass-temperature relation determined from detailed observations
or numerical modeling.
Thus alternatively, as a well-defined observational quantity,
the X-ray temperature function (XTF) has been measured,
which can be converted to the MF by means of the mass-temperature relation.

The cluster temperature function is defined as:
\begin{equation}
N(T,z)=N(M,z) \frac{d M}{dT}
\label{eq:temppp}
\end{equation}
While the mass function, $N(M,z)$, gives the mass and distribution of a population of
evolving clusters, the Jacobian $\frac{d M}{dT}$ describes the physical properties of the single cluster.

Comparison of the predictions of the PS theory with the SCDM and OCDM cosmologies, performed by
Tozzi \& Governato (1998) and Governato et al. (1999), have shown discrepancies between PS predictions and
N-body simulations, increasing with increasing $z$.
\begin{figure}
\label{Fig. 1} \psfig{file=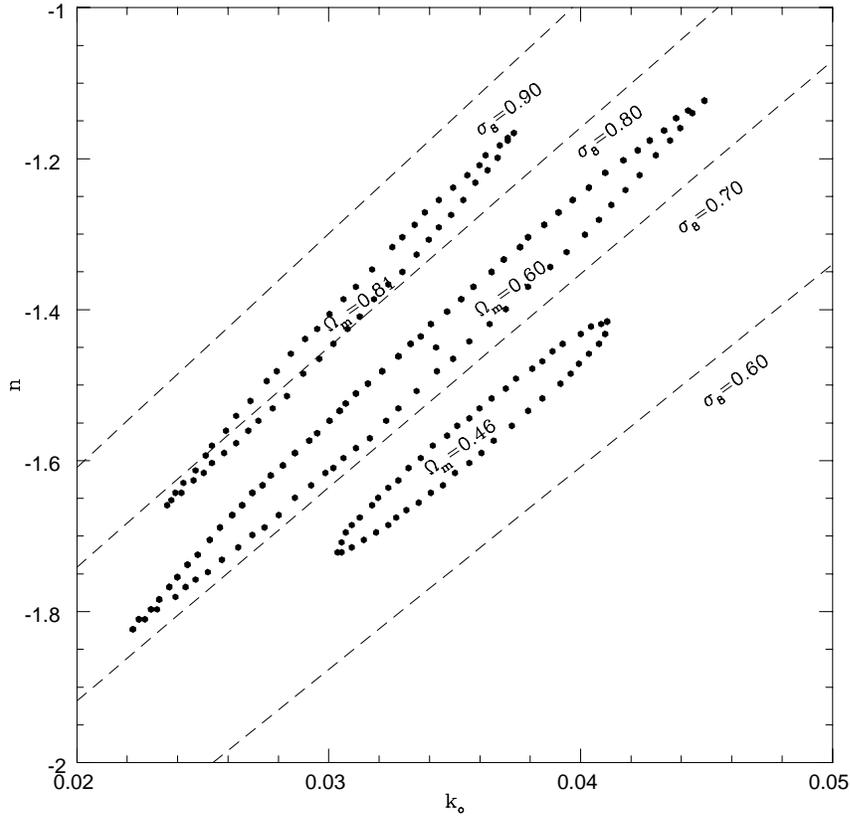,width=12cm}
%\resizebox{12cm}{!}{\includegraphics{fellis1.ps}}
%\parbox[b]{9cm}{
\caption{The $68\%$ confidence contours for the parameters $n$, $k_{\rm o}$ and $\Omega_{\rm m}$ for the open model. The dashed lines are lines of constant $\sigma_8$.
}
%}
\end{figure}
\begin{figure}
\label{Fig. 1}
\psfig{file=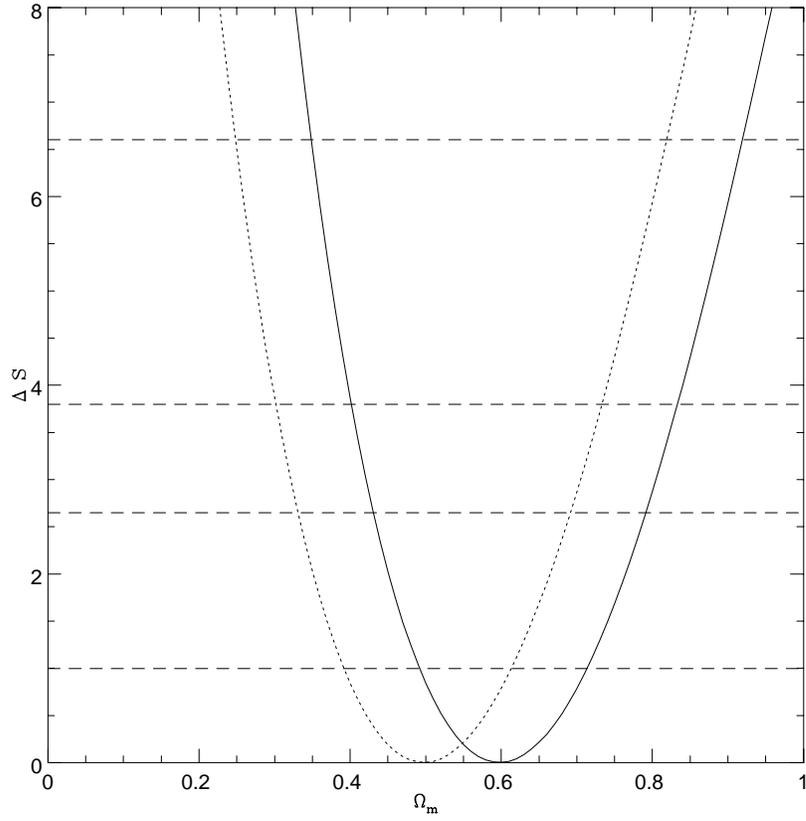,width=12cm}
%\resizebox{12cm}{!}{\includegraphics{fgenry20.ps}}
%\parbox[b]{9cm}{
\caption{ $\Delta$(likelihood)
%contours as a function of
for
the parameter $\Omega_{\rm m}$. The solid line is obtained from the model of this paper while the dotted line is that calculated by Henry (2000). The dashed lines represent various confidence levels (65\%, 90\%, 95\%, 99\%).
}
%}
\end{figure}
\begin{figure}
\label{Fig. 1}
\psfig{file=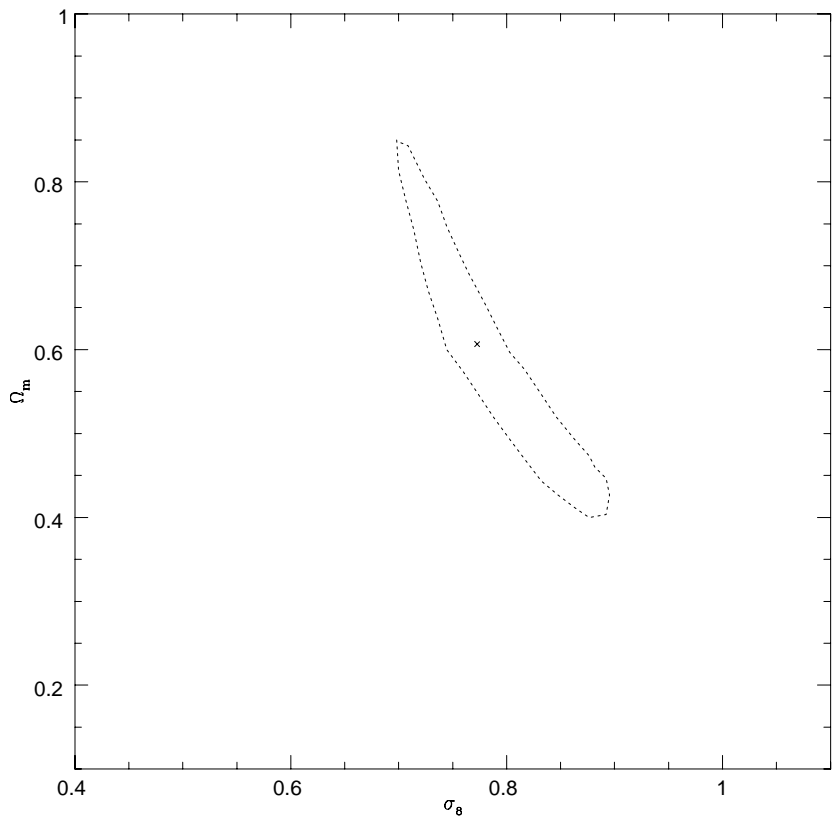,width=12cm}
%\resizebox{12cm}{!}{\includegraphics{omsig1.ps}}
%\parbox[b]{9cm}{
\caption{
The $68\%$ confidence contours for the parameters $\sigma_8$, and $\Omega_{\rm m}$ for the open model (see also Henry (2000), Fig. 9).
%The dashed line is the plot $\Omega_{\rm m}$-$\sigma_8$ of Pierpaoli et al. (2001).
%CORREGGERE: spostare verso destra.
}
%}
\end{figure}
\begin{figure}
\label{Fig. 1}
\psfig{file=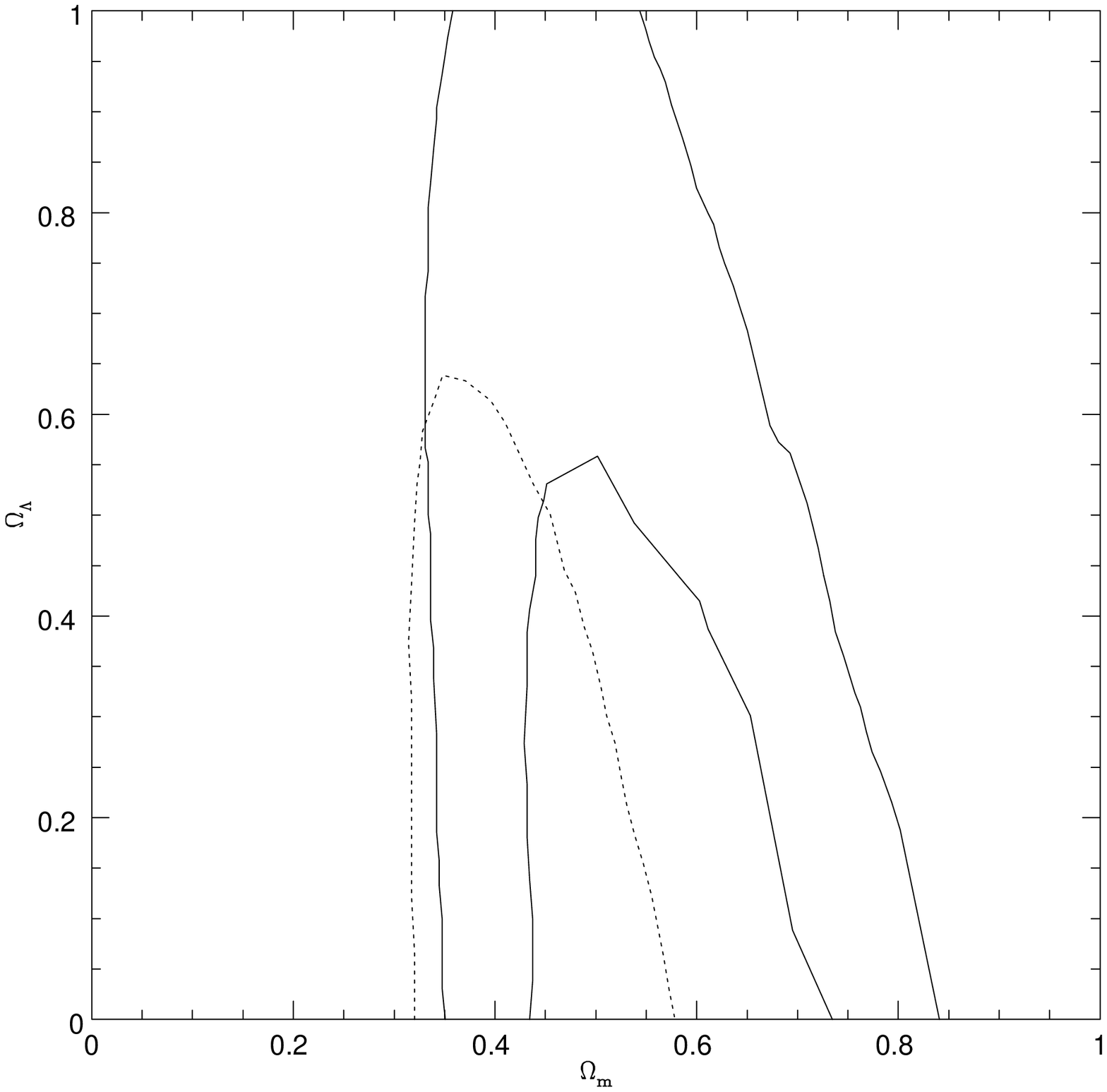,width=12cm}
%\resizebox{12cm}{!}{\includegraphics{fghenry1.ps}}
%\parbox[b]{9cm}{
\caption{
Constraints on $\Omega_{\Lambda}$ and $\Omega_{\rm m}$ obtained using the same 25 clusters used in Henry (2000), for the local sample, while the high redshift sample is constituted from all the EMSS clusters with $z>0.3$ and RX J0152.7-1357
(see Henry 2002). The solid lines are the 1 and 2 $\sigma$ contours obtained using the mass function and the M-T relation of this paper, while the dashed line is the 1 $\sigma$ contour obtained using the PS mass function and the M-T relation of Pierpaoli et al. 2001.
%Nota
%CORREGGERE LA FIGURA
}
%}
\end{figure}

Before going on, I want to discuss about the use of Eq. (\ref{eq:nmm}).
Jenkins et al. (2001), obtained a mass function, which is regarded as perhaps the
most accurate model to date, from the Hubble volume simulations. 
%
%%A reference to an improvement compared to
%%the extended Press-Schechter mass function though useful was important
%%a few years ago, and could appear to be less relevant at the present
%%time. This item could be argumented.
%
\begin{figure}[tbp]
\psfig{figure=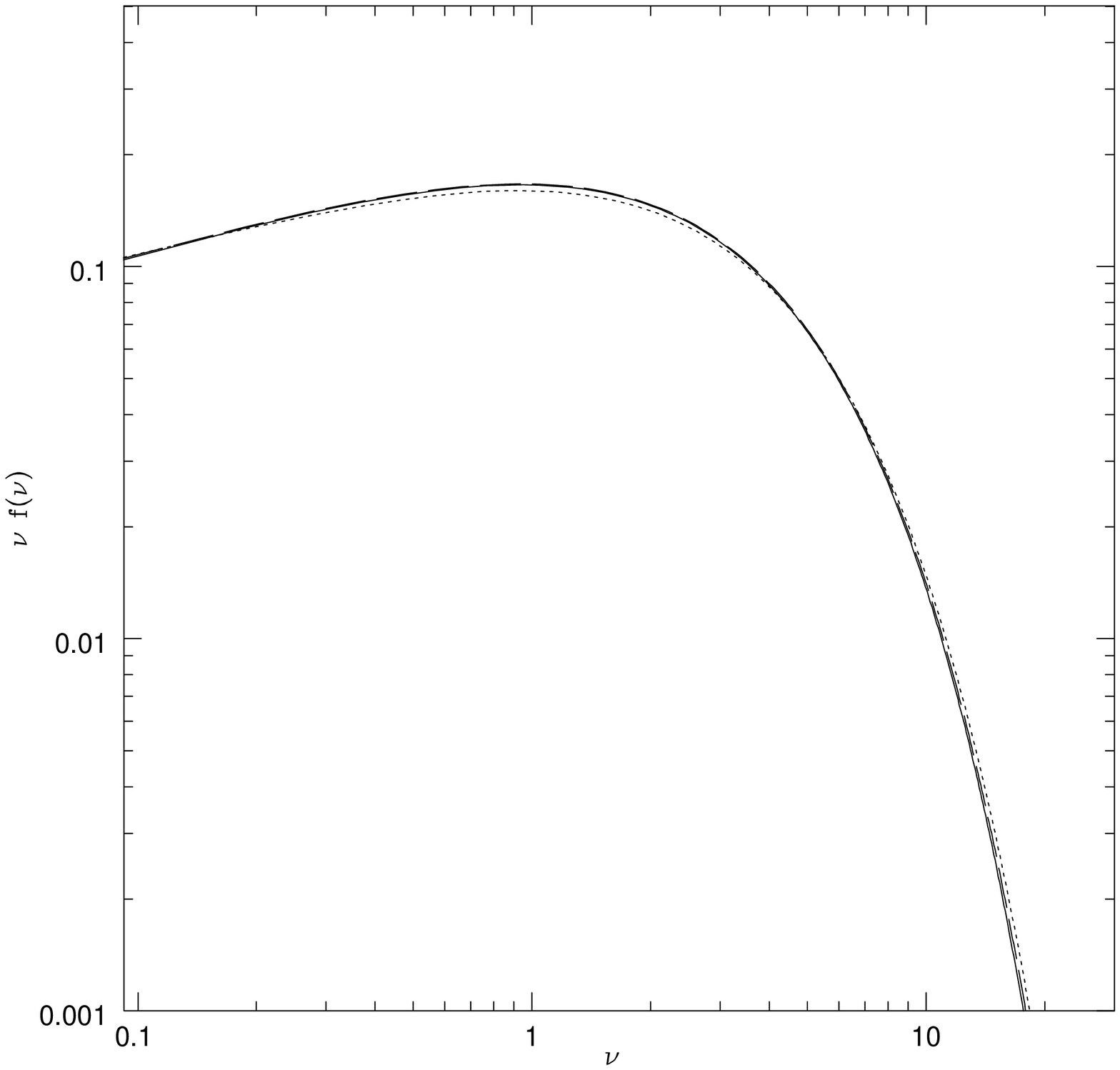,width=12cm} 
\caption{(a) Comparison of various mass functions. The dotted line represents Sheth \& Tormen (2002) prediction, the solid line that of Jenkins et al. (2001) and the dashed line that of Del Popolo (2000b).}
\end{figure}

\begin{figure}
\label{Fig. 1}
\psfig{figure=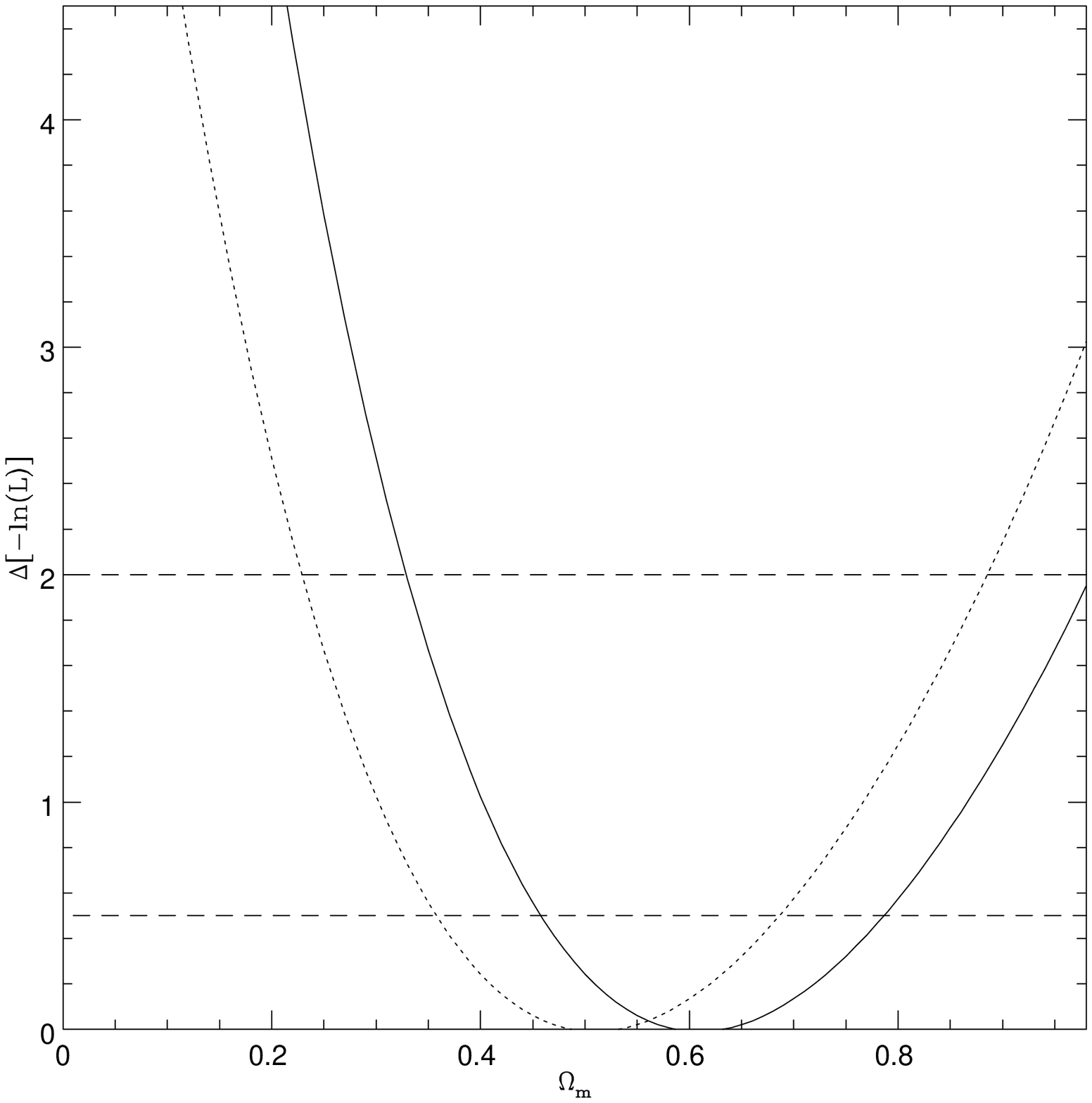,width=12cm} 
\psfig{figure=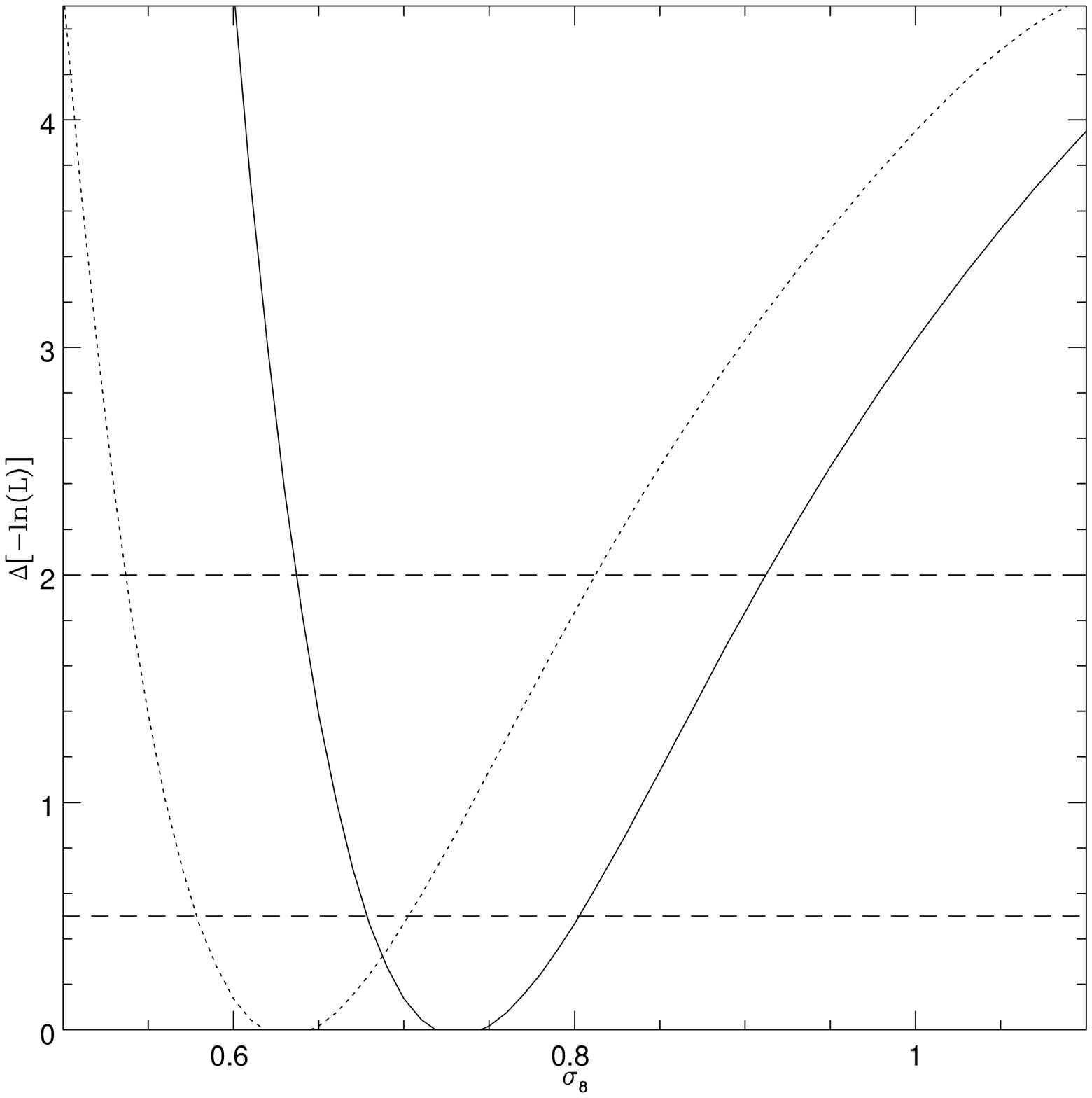,width=12cm} 
%%\resizebox{8cm}{!}{\includegraphics{figeke.ps}}
%%\resizebox{8cm}{!}{\includegraphics{figeke1.ps}}
%&
%\parbox[b]{9cm}{
\caption{
$\Delta(-{\rm ln likelihood})$ for $\Omega_{\rm m}$ (left panel) and $\sigma_8$ (right panel), marginalized over the other two parameters. Solid lines are the prediction obtained from the model of this paper while the dotted lines those obtained from Eke et al. (1998).
Dashed lines are 1 and 2 $\sigma$ significance and the 3 $\sigma$ corresponds to the top of each panel.
}
\end{figure}

Jenkins et al. (2001) showed that, although the mass functions in their 
simulations scaled in accordance with the excursion set prediction, 
Sheth \& Tormen 2002 (Eq. (2)) slightly overestimated the unconditional mass functions 
in their simulations. It interesting to note that as shown by Pierpaoli et al. (2001), using Sheth \& Tormen (2002) or Jenkins et al. (2001, 2003) mass function, rather than PS, the value of parameters like $\sigma_8$ are changed by a small amount (4-8 \%).
Sheth \& Tormen (2002) also showed that 
changing the parameter $a$ in their Eq. (2) from 0.707 to 0.75 reduces the 
discrepancy between it and the simulations substantially. 

In Del Popolo (2002a), I showed that Eq. (\ref{eq:nmm}) of this paper, in agreement with Jenkins et al. (2001), predicts smaller values
of the mass function expecially for high $\nu$, with respect with Sheth \& Tormen's predictions.
In other terms, Eq. (\ref{eq:nmm}) of this paper is in very good agreement with Jenkins et al. (2001). In Fig. 8 I plot a comparison of the various mass functions: the dotted line represents Sheth \& Tormen (2002) prediction, the solid line that of Jenkins et al. (2001) and the dashed line that of Del Popolo (2000b). As shown, Jenkins et al. (2001) mass function is almost indistinguishable from that used in this paper. 
For what concerns small masses, our formula, in agreement with Sheth \& Tormen 2002, differs dramatically from the one Jenkins et al. (2001), propose. 
Simulations currently available do not probe the regime where $\nu \leq 0.3$ or so (the Jeans mass is at 
about $\nu \leq 0.03$ (Sheth \& Tormen 2002)). 
%We hope that simulations in the near future will be able to address which low mass behavior is correct. 
New simulations are needed to address which low mass behavior is correct.
In other terms, the mass function obtained in this paper is in very good agreement with Jenkins et al. (2001) in 
the regime probed by simulations. 
Moreover, the constraints obtained in almost all the papers in literature used the PS mass function except a few papers (e.g. Borgani et al. 2001; Henry 2002) and the M-T relation the usual one obtained from the virial theorem. In other words,
this paper introduces noteworthy improvements on the previous calculations in literature.
The mass entering in Eq. (\ref{eq:nmm}) should be interpreted as the mass contained inside a radius, $r_{180}$ encompassing a mean overdensity $\rho= 180 \overline{\rho}$. However, scaling relations connecting mass X-ray observable quantities may provide the mass at different values of $\rho/\overline{\rho}$. In this case we follow White (2001) and rescale the masses assuming an NFW (Navarro, Frenk \& White 1996) profile for the dark matter halo with a concentration $c=5$ appropriate for rich clusters (see also Pierpaoli et al. 2001, 2003; Schuecher et al. 2002 for more details).

Similarly, it necessary to give arguments to use Eq. (\ref{eq:kTT1}) instead of
the new mass/X-ray temperature relations obtained from simulations or
Chandra data within the last year (see, e.g., Pierpaoli et al. 2001, 2003
for a reference). As shown in Del Popolo (2002a), Eq. (\ref{eq:kTT1}) reduces to a similar equation to that 
used in Pierpaoli et al. (2001) (Eq. 13), in the early-time, namely:
\begin{equation}
M \propto T^{3/2} \rho^{-1/2} \Delta_{\rm c}^{-1/2} \propto T^{3/2} (\Delta_{\rm c} E^2)^{-1/2}
\label{eq:pierp}
\end{equation}
where $E(z)^2=\Omega_{\rm M} (1+z)^3+\Omega_{\Lambda}$, 
and the term depending on $\Omega_{\Lambda}$ in Eq. (13) of Pierpaoli et al. (2001) (which is a correction to the 
virial relation arising from the additional $r^2$ potential in the presence of $\Lambda$) is neglected since it produces only a small correction (see Pierpaoli et al. 2001). 
I also want to add that Eq. (13) of Pierpaoli et al. (2001) or Eq. (4) of Pierpaoli et al. (2003), 
%which is Eq. (\ref{eq:pierp}), 
comes from rather simplistic arguments (dimensional analysis and an assumption that clusters are self-similar) 
and is a good approximation to both observations and simulations
%is a remarkable good fit to the simulations, 
but this last are 
sufficiently computationally demanding that they cannot explore parameter space efficiently and so it is necessary to determine coefficients by means of simulations, while scalings are taken from simple theoretical models (Pierpaoli et al. 2001). 
%Eq. (\ref{eq:kTT1}) does not have these drawbacks.????????( TOGLIERE)
Eq. (13) of Pierpaoli et al. (2001), is valid for systems hotter than about 3 keV.
We know that recent studies have shown that the self-similarity in the M-T relation seems to break at 
some keV (Nevalanien et al. 2000; Xu, Jin \& Wu 2001): Afshordi \& Cen (2001) has shown that non-sphericity introduces an asymmetric, mass dependent, scatter for the M-T relation altering its slope at the low mass end ($T \sim 3$ keV). These effects are taken into account by mine M-T relation, which as previously told gives same results of that of
Pierpaoli et al. (2001, 2003) in the range of energy in which this is valid.

In the following, I'll use the mass function modified as
described in the previous section
%in Eq. (\ref{eq:nmm})
and an
improved form of the M-T relation in order to calculate the mass function.

Introducing Eq. (\ref{eq:nmm}) into Eq. (\ref{eq:temppp}) and using the M-T relation, in the peculiar case that
the variance is given by:
\begin{equation}
\sigma=C (\frac{M}{10^{15} M_{\odot}})^{-(3+n)/6}=C m^{-(3+n)/6}
\end{equation}

I get:
~\\
~\\

%%\[
%%\left\{
%\begin{array}{lc}

\begin{equation}
\frac{dN}{dT}=\frac{N_1(m,z)}{N_2(m,z)}
\end{equation}

\begin{eqnarray*}
N_1(m,z) &=&1/2\,{ m_1}\,\left (3+n\right )\rho\,\delta\,\sqrt {a}\sqrt
{2}A{m}^{{ m_2}+q}
\nonumber \\
& &
{e^{-\frac{1}{2C^2}\,{m}^{{ m_3}}a{\delta}^{2}-{a}^{1-
\alpha}{\delta}^{2-2\,\alpha}{m}^{{ m_3}}\left (C {m}^{{ m_4}}
\right )^{2\,\alpha}{ a_2}-1/2\,{a}^{1-2\,\alpha}{\delta}^{2-4\,
\alpha}{m}^{{ m_3}}\left (C {m}^{{ m_4}}\right )^{4\,\alpha}{{ a_2
}}^{2}}}
\nonumber \\
& &
\left (1+{ a_1}\,{a}^{-\alpha}{\delta}^{-2\,\alpha}\left ({C m
}^{{ m_4}}\right )^{2\,\alpha}\right )
\end{eqnarray*}

\begin{eqnarray*}
N_2(m,z) &=& {m}^{7/6}(3\,F{{ m_1}}^{2}{ LPh}\,q{m}^{{ a_3}}-3\,F{
m_1}\,{ a_3}\,{ m_{1l}}\,{ LPh}\,{m}^{{ a_3}}-2\,F{ LPh}\,{{
m_1}}^{2}{m}^{{ a_3}}+
\\
& &
3\,F{{ m_1}}^{2}{ a_3}\,{ m_{1l}}\,{
LPh}\,{m}^{{ a_3}}+2\,F{ LPh}\,{ m_1}\,{m}^{{ a_3}}
-3\,F{
m_1}\,{ LPh}\,q{m}^{{ a_3}}-3\,F{{ m_1}}^{2}{ a_3}\,{
LPh}\,{m}^{{ a_3}}+
\\
& & 
3\,F{ m_1}\,{ a_3}\,{ LPh}\,{m}^{{ a_3}}
-3\,F{{ m_1}}^{2}{ LPh}\,q-3\,F{ m_1}\,{ a_3}+2\,F{ LPh}\,
{{ m_1}}^{2}-2\,{m}^{q}
-2\,{ t_1}\,{ m_1}\,{m}^{q}-
\\
& &
3\,F{{ m_1}
}^{2}{ a_3}\,{ m_{1l}}\,{ LPh}+3\,F{ m_1}\,{ a_3}\,{ m_{1l}}
\,{ LPh}+3\,F{{ m_1}}^{2}{ a_3}-2\,F{ LPh}\,{ m_1}+3\,F{
m_1}\,{ LPh}\,q){ C}\,\sqrt {\pi}
\end{eqnarray*}

%\begin{equation}
\[
kT \simeq 8 keV m^{2/3}
\frac{
\left[
\frac{1}{m_1}+t_1^{\frac 23}
+\frac{K(m_1,x)}{(M_{15} m)^{8/3}}
\right]
}
{
\left[
\frac{1}{m_1}+t_1^{\frac 23}
 +\frac{K_0(m_1,x)}{M_{0}^{8/3}}
\right]
}
\]
%\end{equation}
%%%%%%%%%%%%%%%%\end{array}
%%\right.
%%\]

%
%%NOTE
%%a) RIVEDERE $M_8$ IN kT...(legame tra $M_{15}$ ed $M_{8}$ cambiamenti in kT. Nelle pagine precedenti kT contiene $M_0$ che %%legame c'e'?)
%
%%b) RIVEDERE I TERMINI MANCANTI $\Omega_M$ ALL'INZIO DELLA PS, $D_0$, $D_z$
%%c) Verificare la formula di Henry
%

In order to use the same notation and variance of Henry (2000), the constant $C$ is defined as:
\begin{equation}
C=0.675\,\sqrt {{\frac {{\Gamma}(3+n)\sin(1/2\,n\pi )}{{2}^{n}n
\left (2+n\right )\left (1-n\right )\left (3-n\right )}}}\left (
 857.375\,{\frac {{k_{{o}}}^{3}{{\it Mpc}}^{3}}{{h}^{2}\Omega_{\rm m}}}\right
)^{-1/2-1/6\,n}
\end{equation}
and then $\sigma_8=\sigma(\Omega_{\rm m}, M=0.594 \times 10^{15} h^{-1} \Omega_{\rm m})$, and the XTF depends on the
parameters $n$, $k_{\rm o}$ and $\Omega_{\rm m}$.
The data that shall be fitted to the theory previously described, are those described in Section. 2 of Henry (2000).
I use a maximum likelihood fit to the unbinned data in order to determine various model parameters.
The method is described in Marshall et al (1983). The likelihood function is given by their Eq. (2),
which for the present situation is:
%I use the maximum likelihood method with the likelihood function:
\begin{equation}
S=-2\sum_{i=1}^{N}\ln \left[ n(\Omega _{m},z_{i},kT_{i})\frac{d^{2}V(\Omega _{m},z_{i})}{dzd\Omega }\right] +2\int_{kT_{\min }}^{kT_{\max }}dkT\int_{z_{\min }}^{z_{\max }}n(\Omega _{m},z,kT)\Omega (z,kT)\frac{d^{2}V(\Omega _{m},z)}{dzd\Omega }dz
\label{eq:likelih}
\end{equation}
(Henry 2000),
where N is the number of clusters observed, $n(\Omega_{\rm m},z,kT)$ is the temperature function, $\Omega(z,kT)$ is the solid
angle in which a cluster with temperature kT at redshift z could have been detected (the selection
function) and $\frac{d^{2}V(\Omega _{m},z_{i})}{dzd\Omega }$ is the differential volume, which is given in the Appendix of Henry (2000) (see also Henry 2000 for a description of $\frac{d^{2}V(\Omega _{m},z_{i})}{dzd\Omega }$ in the first term of Eq. (\ref{eq:likelih})).
The best estimates for the parameter are obtained minimizing $S$ and the confidence regions are obtained as in Henry (2000) by noticing that $S$ is distributed as $\chi^2$ with the number of degree of freedom equal to the number of interesting parameters (Avni 1976). The model is accepted or rejected according to what prescribed in Marshall et al. (1983) (see also Henry 2000).

At this point, we can fit the data described in Section. 2 of Henry (2000) to the theory previously described using
the quoted maximum likelihood method.
%The best fitting integral temperature functions are shown in
%Figures 1 and 2. The two cumulants given in equations (7) and (8) are uniform with a probability of 0.70
%and 0.14 respectively for the open model and 0.78 and 0.13 respectively for the flat model. Thus each fit
%is acceptable.
The most general description of the results requires the three parameters of the fit. I show these results
in Fig. 4, where I plotted the results for the open model.
It is straightforward to read off the value of n, which is
at the 68\% confidence for the open model.
%and flat models respectively.
These values shows that the correction introduced by the new form of the mass function and M-T relation gives rise to higher values of $\Omega_{\rm m}$ ($\Omega_{\rm m}=0.6 \pm 0.13$, while it is $\Omega_{\rm m}=0.49 \pm 0.12$ for Henry (2000)) and
$n=-1.5 \pm 0.32$ ($n=-1.72 \pm 0.34$ in Henry (2000)).
%
%%The shape parameter that we adopt, n, may be converted to the shape
%%parameter of the popular cold dark matter fluctuation spectrum, $\Gamma$, using Eq. (5) of Pen (1998) or
%%Eqs. (2) and (3) of Viana and Liddle (1999). Both give nearly the same results, which are that our fit
%%implies $\Gamma \simeq 0.05 ± 0.22$ FARE IL CALCOLO
%
The presentation in Fig. 4 is somewhat difficult to appreciate,
so we also give the constraints for
fewer parameters.
In Fig. 5, I plot
%$\Omega_{\rm m}$.
$\Delta$(likelihood)
%contours as a function of
for the parameter $\Omega_{\rm m}$. The solid line is obtained from the model of this
paper while the dotted line is that calculated by Henry (2000). The dashed lines represent various confidence levels (68\%, 90\%, 95\%, 99\%).
Constraints are relatively tight when
considering this single parameter. We find that $\Omega_{\rm m}=0.6^{+0.12}_{-0.11}$
at the 68\% confidence level and at $\Omega_{\rm m}=0.6^{+0.23}_{-0.2}$ the
at 95\% confidence level for the open model.

%For the flat model the corresponding values are W 0 = 0.44 12
%at the 95\% confidence level. These results agree with those in H97,
%except the errors are ~ 40\% smaller here.
%
%%?????A closed universe is ruled out at much greater than 99\% confidence.
%
%We recast
The constraints in Fig. 4 are plotted into a more conventional format in Fig. 6. Three
parameters are still required, but the constraints on n and $k_0$ are collapsed into $\sigma_8$ .
%Some of the degeneracy
%between these two parameters still remains in the form of a banana shaped contour particularly for the flat
%model. Still, we are able to make a relatively precise measurement of s 8 , finding values of 11
%at 68\% confidence level for the open and flat cases respectively. These results also agree with
%those in H97.
Fig. 6, plots the $68\%$ confidence contours for the parameters $\sigma_8$, and $\Omega_{\rm m}$ for the open model (see also Henry (2000), Fig. 9).
%The dashed line is the plot $\Omega_{\rm m}$-$\sigma_8$ of Pierpaoli et al. (2001).
In Fig. 7, I plot the
constraints on $\Omega_{\Lambda}$ and $\Omega_{\rm m}$ obtained using the same 25 clusters used in Henry (2000), for the local sample, while the high redshift sample is constituted from all the EMSS clusters with $z>0.3$ and RX J0152.7-1357
(see Henry 2002). The solid lines are the 1 and 2 $\sigma$ contours obtained using the mass function and the M-T relation of this paper, while the dashed line is the 1 $\sigma$ contour obtained using the PS mass function and the M-T relation of Pierpaoli et al. 2001.
%Nota
%PER $\sigma_8$ SEMBRA CHE EKE et al 1998 e BORGANI diano risultati opposti.........

For a CDM spectrum, the expression for the XTF is much more complicated. It can be obtained combining Eq. (\ref{eq:nmm}),
Eq. (\ref{eq:temppp}), and our M-T relation.
The mass variance
can be obtained once a spectrum, $P(k)$, is fixed, by:
\begin{equation}
\sigma ^2(M)=\frac 1{2\pi ^2}\int_0^\infty dkk^2P(k)W^2(kR)
\label{eq:ma3}
\end{equation}
where $W(kR)$ is a top-hat smoothing function:
\begin{equation}
W(kR)=\frac 3{\left( kR\right) ^3}\left( \sin kR-kR\cos kR\right)
\label{eq:ma4}
\end{equation}
and the power spectrum $P(k)=Ak^nT^2(k)$ is fixed giving the transfer
function $T(k)$.
The CDM spectrum used in this paper is that of Bardeen et al. (1986)(equation~(G3)), with transfer function:
\begin{equation}
T(k) = \frac{[\ln \left( 1+2.34 q\right)]}{2.34 q}
\cdot [1+3.89q+
(16.1 q)^2+(5.46 q)^3+(6.71)^4]^{-1/4}
%
%T^2(k) &=& [\ln \left( 1+4.164k\right)]^2 \cdot (192.9+1340k+ \nonumber \\
%& + &  1.599\cdot 10^5k^2+1.78\cdot 10^5k^3+3.995\cdot
%10^6k^4)^{-1/2}
%
\label{eq:ma5}
\end{equation}
Bardeen et al. (1986)(Eq. ~(G3)),
where
%$ A$ is the normalizing constant and
$q=\frac{k\theta^{1/2}}{\Omega_{\rm X} h^2 {\rm Mpc^{-1}}}$.
Here $\theta=\rho_{\rm er}/(1.68 \rho_{\rm \gamma})$
represents the ratio of the energy density in relativistic particles to
that in photons ($\theta=1$ corresponds to photons and three flavors of
relativistic neutrinos).
Using the data used by Eke et al. (1998) (see their Sec. 3 \footnote{They combined the temperature data for 25 local clusters by Henry \& Arnaud (1991) with the sample of 10 {\it EMSS} clusters at $0.3<z<0.4$ by Henry et al. (1997)}) and the a maximum likelihood parameter estimation (see Eke et al. 1998, Sec. 4):
\begin{equation}
\ln L=\sum_{i=1}^{N}a_{i}-\int \int adzdT
\end{equation}
where
\begin{equation}
a(z,t)=n(z,T,p)\frac{dV(z,p)}{dz}\varsigma (z,t,p)
\end{equation}
where $p$ represents the three model parameters being investigated, $n(z,T,p)$
is the comoving cluster number density given by Del Popolo (2000b) model for the mass function,
$V$ is the comoving volume
and $\varsigma$ is the product of the fraction of the sky surveyed and the estimated completeness of the survey (see Eke et al. (1998), Sec. 4, for details).
%
%%NOTA:
%%A COSA SERVE LA L-M????? COME SI CALCOLA LA $\varsigma (z,t,p)$????????? CHIEDERE a EKE
%
%%CORREGGERE LE FIGURE CON $\sigma_8$, maggiore o minore?
%%..................
%

The results are plotted in Fig. 9. It plots the
$\Delta(-{\rm ln likelihood})$ for $\Omega_{\rm m}$ (left panel) and $\sigma_8$ (right panel), marginalized over the other two parameters. Solid lines are the prediction obtained from the model of this paper while the dotted lines those obtained from Eke et al. (1998).
Dashed lines are 1 and 2 $\sigma$ significance and the 3 $\sigma$ corresponds to the top of each panel.
Fig. 9 shows that $\Omega_{\rm m}$ (left panel) and $\sigma_8$ (right panel) are increased with respect to Eke et al. (1998) prediction:
while in Eke et al. (1998) $\Omega_{\rm m}=0.52^{+0.17}_{-0.16}$, and
$\sigma_8= 0.63^{+0.08}_{-0.05}$,
I find that $\Omega_{\rm m}= 0.62^{+0.17}_{-0.15}$
and
$\sigma_8= 0.73^{+0.07}_{-0.06}$. This shows again an increase in $\Omega_{\rm m}$, also in qualitative agreement with
Eke et al. (1998) calculation taking account of changes in the threshold for collapse suggested by Tozzi \& Governato (1998).
%according to Eke et al. (1998) $\Omega_{\rm m}=0.52^{+0.17}_{-0.16}$ and $\sigma_8=0.63^{+0.08}_{-0.05}$ while I find %$\Omega_{\rm m}=0.52^{+0.17}_{-0.16}$ and $\sigma_8=0.63^{+0.08}_{-0.05}$

\section{Results and discussion}
%\section{Comparison with previous studies}
In this paper, I have revisited the constraints obtained
by several authors (Reichart et al. 1999; Eke et al. 1998; Henry 2000)
on the estimated values of $\Omega_{\rm m}$, $n$ and $\sigma_8$ in the light of recent theoretical developments: new theoretical mass functions, a more accurate mass-temperature relation, also determined for arbitrary $\Omega_{\rm m}$ and $\Omega_{\rm \Lambda}$.
Using the mass function derived in Del Popolo (2002b), the M-T relation derived in Del Popolo (2002a), and following Reichart et al. (1999), I calculated the XLF which can be applied to the high-redshift X-ray cluster luminosity catalogs to constraints cosmological parameters, namely in this case $\Omega_{\rm m}$ and $n$. This luminosity function was applied, for a fixed value of $H_0=50 {\rm km Mpc^{-1} s^{-1}}$, to broad subsets of the revised EMSS X-ray cluster subsample of Nichol et al. (1997) and to ROSAT BCS luminosity function of Ebeling et al. (1997) to constraint $\Omega_{\rm m}$.
For the 61 revised EMSS clusters, with $0.14<z<0.6$, I find that
$\Omega_{\rm m}=1.15^{+0.40}_{-0.33}$ and
$n=-1.55^{+0.42}_{-0.41}$.
The previous result shows that the change in the mass function and M-T relation gives rise to an increase of $\Omega_{\rm m}$ and $n$ of $\simeq 20\%$. Then, taking account of the fact that
massive clusters accrete matter quasi-continuously and taking account of non-sphericity in collapse changes the values of the estimated parameters. The principal interest in this paper is to study the effects of improvements on the mass function and M-T relation on the values of cosmological parameters, and not in the peculiar value obtained.
So, I have also estimated the value of $\Omega_{\rm m }$
following Borgani et al. (2001). In their study, they used as mass function the ST mass function instead of the usual PS,
(differently from Reichart et al. (1999)), which is a good approximation to the function I used in this paper. They used
Eke et al.(1998) M-T relation, which is different from the one I used..
Using their method and data (RDCS), but our mass function and M-T relation, I obtained a larger value of $\Omega_{\rm m}$ ($\Omega_{\rm m}= 0.4 \pm 0.1$ instead of $\Omega_{\rm m}=0.35^{+0.13}_{-0.10}$) that differently from the previous analysis (Reichart et al. 1999) exclude an Einstein-de Sitter model, but shows again that a change in the mass function and M-T relation influences the value of the parameters to constraint.
Another possibility to constraint cosmological parameters is to use the XTF.
I have repeated the Henry (2000) analysis but differently from the quoted paper, I changed the mass function and M-T relation, adopting again those of Del Popolo (2000a,b). The qualitative result is similar to the previous one, using the XLF, namely the values of the obtained cosmological parameters are modified by a $ \simeq 20\%$.
The new form of the mass function and M-T relation gives rise to higher values of $\Omega_{\rm m}$ ($\Omega_{\rm m}= 0.6 \pm 0.13$ in my estimation, while it is $\Omega_{\rm m}= 0.49 \pm 0.12$ for Henry (2000)) and
$n=-1.5 \pm 0.32$, in my estimation, while $n=-1.72 \pm 0.34$ in Henry (2000).
I have also obtained some constraints on $\Omega_{\Lambda}$ and $\Omega_{\rm m}$ obtained using the same 25 clusters used in Henry (2000), for the local sample, while the high redshift sample is constituted from all the EMSS clusters with $z>0.3$ and RX J0152.7-1357
(see Henry 2002). The 1 $\sigma$ contours obtained using the mass function and the M-T relation of this paper, and plotted in Fig. 7, shows that for $\Lambda=0$,
it is $0.32<\Omega_{\rm m}<0.57$ in the case of Henry (2002) and $0.43<\Omega_{\rm m}<0.73$ in my estimation. The figure shows the constraints to $\Omega_{\rm m}$ for different values of $\Omega_{\rm \Lambda}$.
Similar results to that obtained in the previous comparison with the Henry (2000) results
are obtained changing data and method, by following Eke et al. (1998). I obtain a value of $\Omega_{\rm m}= 0.62^{+0.17}_{-0.15}$, while in Eke et al. (1998) $\Omega_{\rm m}= 0.52^{+0.17}_{0.16}$, and $\sigma_8= 0.73^{+0.07}_{-0.06}$ while in Eke et al. (1998) $\sigma_8= 0.63^{+0.08}_{-0.05}$. This shows again an increase in $\Omega_{\rm m}$, also in agreement with Eke et al. (1998) calculation taking account of changes in the threshold for collapse suggested by Tozzi \& Governato (1998). \footnote{Replacing $\delta_{\rm c}$ with the $\delta_{\rm eff}$ of Tozzi \& Governato (1998), produces qualitatively similar changes on the mass function as those obtained using the Del Popolo (2002b) mass function.}
As previously told, this paper has its aim that of studying how ``systematic uncertainties" can influence the values of some cosmological parameters. It is well known that in literature the values obtained for $\Omega_{\rm m}$ span the range $ 0.2 \leq \Omega_{\rm m} \leq 1$ (Reichart et al. 1999).

Sadat, Blanchard \& Oukbir (1998) and Reichart et al. (1999)
Blanchard \& Bartlett (1998)
found results consistent with $\Omega_{\rm m}=1$. Viana \& Liddle (1999) found that
$\Omega_{\rm m}= 0.75$ with $\Omega_{\rm m}>0.3$ at the 90\% confidence level and $\Omega_{\rm m} \simeq 1$ still viable. Blanchard, Bartlett \& Sadat (1998) found almost identical results ($\Omega_{\rm m} \simeq 0.74$ with $0.3<\Omega_{\rm m}<1.2$ at the 95\% confidence level).
Eke et al. (1998) found $\Omega_{\rm m}=0.45 \pm 0.2$. It is interesting to note (as previously mentioned) that Viana \& Liddle (1999) used the same data set as Eke et al. (1998) and showed that uncertainties both in fitting local data and in the theoretical modeling could significantly change the final results: they found $\Omega_{\rm m} \simeq 0.75$ as a preferred value with a critical density model acceptable at $<90\%$ c.l.

Different results were obtained by Bahcall, Fan \& Cen (1997) ($\Omega_{\rm m}=0.3 \pm 0.1$), Fan, Bahcall \& Cen (1997) ($\Omega_{\rm m}= 0.3 \pm 0.1$), Bahcall \& Fan (1998) ($\Omega_{\rm m}=0.2^{+0.3}_{-0.1}$) and several other authors.
%So summarizing, the value of $\Omega_{\rm m}$ obtained changes according to the method and data used and sometime the same %data can lead........\\

The reasons leading to the quoted discrepancies has been studied in several papers (Eke et al. 1998; Reichart et al. 1999; Donahue \& Voit 1999; Borgani et al. 2001). According to Reichart (1999) unknown systematic effects may be plaguing great part of the quoted results. A list of these last is reported in the introduction to this paper.
Our analysis shows that improvements in the mass function and M-T relation increases the value of $\Omega_{\rm m}$. The effect of this increase is unable to enhance significantly the probability that $\Omega_{\rm m}=1$ in the case of
constraints like that of Fan, Bahcall \& Cen (1997) ($\Omega_{\rm m}=0.3$) or Bahcall \& Fan (1998) ($\Omega_{\rm m}=0.2$), and can give a small contribution even in the case of larger values for the value of the constraints obtained.
However, in any case it shows that even small correction in the physics of the collapse can induce noteworthy effects on the constraints obtained. Moreover, even changing the data or the way they are analyzed gives different results. As an example, changing their low-redshift sample, Donahue \& Voit (1999) showed a change in $\Omega_{\rm m}$ from 0.45 to 0.3.
%Apart from the theoretical uncertainties, the improvement in observations requires an improvement in the theoretical %framework....
%The art of determining an X-ray cluster catalog's selection function is a constantly improving science: modern selection %functions are determined via extensive numerical simulations
Furthermore, as observations are reaching the first epoch of cluster assembly, treating them as dynamical relaxed and virialized systems is undoubtly an oversemplification. Hierarchical clustering scenario predicts that
a fraction between 0.3 and 0.6 of the $z=1$ population of clusters are observed less than 1 Gyr after the last major merger event and then are likely to be in a state of non-equilibrium. Although the quoted uncertainties has been so far of minor importance with respect to the paucity of observational data, a breakthrough is needed in the quality of the theoretical framework if high-redshift clusters are to take part in the high-precision-era of observational cosmology.

%AGGIUNGERE FIGURA 3 DI REICHART
%E FORSE ANCHE LE 2 di HENRY 2000 di N(>KT)


\begin{thebibliography}{}
\bibitem{} Afshordi N., Cen R., 2001, astro-ph/0105020
\bibitem{} Avni Y., Bahcall J.N., 1980, ApJ 235, 694
\bibitem{} Bahcall N.A., Cen R., 1993, ApJ 407, L49
\bibitem{} Bahcall N.A., Fan X., Cen R., 1997, ApJ 485, L53
\bibitem{} Bahcall N.A., Fan X., 1998, ApJ 504, 1
\bibitem{} Bardeen J.M., Bond J.R., Kaiser N., Szalay A.S., 1986, ApJ 304, 15
\bibitem{} Bialek J.J., Evrard A.E., Mohr J.J., 2001, ApJ 555, 597
\bibitem{} Biviano A., Girardi M., Giuricin, G., Masrdirossian F., Mezzetti M., 1993, ApJ 411, L13
\bibitem{} Blanchard A., Bartlett J.G., 1997, A\&A 332, L49
\bibitem{} Blanchard A., Bartlett J.G., Sadat R., 1998, in Les Comptes Rendus de l'Academie des Sciences
%, in press.....
\bibitem{} Blanchard A., Sadat R., Bartlett J.G., Le Dour M., 2000, A\&A 362, 809
\bibitem{} Bond J.R., Myers S.T., 1991, Trends in Astroparticle Physics, eds. D. Cline, R. Peccei, World scientific, Singapore, p. 262
\bibitem{} Borgani S., Girardi M., Carlberg R. G., Yee H. K. C., Ellingson E., 1999, ApJ 527, 561
\bibitem{} Borgani S., Guzzo L., 2001, Nature 409, 39
\bibitem{} Borgani S., Rosati P., Tozzi P., Stanford S. A., Eisenhardt P. R., Lidman C., Holden B., Della Ceca R.,  Norman C., Squires G., 2001, ApJ 561, 13
\bibitem{} Bryan G.L., Norman M.L., 1997, ApJ 495, 80
\bibitem{} Carlberg R.G., Yee H.K.C., Ellingson E., Abraham R., Gravel p., Morris S.L., Pritchet C.G., 1996, ApJ 462, 32
\bibitem{} Cash A., 1979, ApJ 228, 939
\bibitem{} Colafrancesco S., Mazzotta P., Vittorio N., 1997, ApJ 488, 566
\bibitem{} David L.P., Slyz A., Jones C., Forman W., Vrtilek S.D., 1993, ApJ 412, 479
\bibitem{} Del Popolo A., Gambera M., 1998, A\&A 337, 96
\bibitem{} Del Popolo A., Gambera M., 1999, A\&A 344, 17
\bibitem{} Del Popolo A., Gambera M., 2000, A\&A 357, 809
\bibitem{} Del Popolo, A., E. N. Ercan, Z. Q. Xia, 2001, AJ 122, 487
\bibitem{} Del Popolo A., 2002a, MNRAS 336, 81
\bibitem{} Del Popolo A., 2002b, MNRAS 337, 529
\bibitem{} Donahue M., 1996, ApJ 468, 79
\bibitem{} Donahue M., Gioia I., Luppino G., Hughes J.P., Stocke J.T.,
%1998, ApJ....
(astro-ph/9707010)
\bibitem{} Donahue M., Voit G.M., 1999, astro-ph/9907333
\bibitem{} Ebeling H., Edge A. C., Fabian A. C., Allen S. W., Crawford C. S., Boehringer H., 1997, ApJ 479, 101
\bibitem{} Edge A.C., Stewart G.C., Fabian A.C., Arnaud K.A., 1990, MNRAS 245, 559
\bibitem{} Efstathiou G., Frenk C.S., White S.D.M., Davis M., 1988, MNRAS 235, 715
%\bibitem{} Efstathiou G., Frenk C.S., White S.D.M., Davis M., 1988, MNRAS 235, 715
\bibitem{} Evrard A.E., 1989, ApJ 341, L71
\bibitem{} Eke V.R., Cole S., Frenk C.S., 1996, MNRAS 282, 263
\bibitem{} Eke V.R., Cole S., Frenk C.S., Navarro J.F., 1996, MNRAS 281, 703
\bibitem{} Eke V. R., Cole S, Frenk C. S., Henry J.P., 1998, MNRAS 298, 1145
\bibitem{} Evrard A.E., 1990, in Proc. STScI Symp.4, ed. W.R. Oegerle, M.J. Fitchett, \& L. Danly (New York: Cambridge Univ. Press), 287
\bibitem{} Evrard A.E., Metzler C.A., Navarro J.F., 1996, ApJ 469, 494
\bibitem{} Evrard A.E., 1997, MNRAS 292, 289
\bibitem{} Fan X., Bahcall N.A., Cen R., 1997 ApJ 490, L123
\bibitem{} Finoguenov A., Reiprich T.H., B\"oeringer H., 2001, A\&A 368, 749
\bibitem{} Gioia I.M., Luppino G.A., 1994, ApJS 94, 583
\bibitem{} Girardi M., Borgani S., Giuricin C., Mardirossian F., Mezzetti M., 1998, ApJ 506, 45
\bibitem{} Governato F., Babul A., Quinn T, Tozzi P., Baugh C., Katz N., Lake G., 1999, MNRAS 307, 949
\bibitem{} Gregory P.C., Loredo T., ApJ 398, 146
\bibitem{} Gross M.A.K., Sommerville R.S., Primack J.R., Holtzman J., Klypin A., 1998, MNRAS 301, 81
\bibitem{} Henry J.P., Arnaud K.A., 1991, ApJ 372, 410
\bibitem{} Henry J.P., Gioia I.M., Maccacaro T., Morris S.L., Stocke J.T.,  Wolter A., 1992, ApJ 386, 408
\bibitem{} Henry J.P., 1997, ApJ 489, L1
\bibitem{} Henry J.P., 2000, ApJ 534, 565
\bibitem{} Henry J. P., 2002, ASP Conference Proceedings, Vol. 257. Edited by Lin-Wen Chen, Chung-Pei Ma, Kin-Wang Ng, and Ue-Li Pen.
\bibitem{}Kitayama, T., Suto, Y., 1997, ApJ 490, 557
\bibitem{} Komatsu E., Seljak U., 2001, MNRAS 327, 1353
\bibitem{} Lacey C., Cole S., 1993, MNRAS 262, 627
\bibitem{} Lacey C., Cole S., 1994, MNRAS 271, 676
\bibitem{} Marshall H. L., Avni Y., Tananbaum H., \& Zamorani G. 1983, ApJ 269, 35
\bibitem{} Mathiesen B., Evrard A.E., 1998, MNRAS 295, 769
\bibitem{} Muanwong O., Thomas P. A., Kay S. T., Pearce F. R., Couchman H. M. P., 2001, ApJ 552, 27, (astro-ph/0102048)
\bibitem{} Navarro J.F., Frenk C.S., White S.D.M., 1995, MNRAS 275, 720
\bibitem{} Nevalainen J., Markevitch M., Forman W., 2000, ApJ 532, 694
\bibitem{} Nichol R. C., Holden B. P., Romer A. K., Ulmer M. P., Burke D. J., \& Collins, C. A. 1997, ApJ 644
\bibitem{} Oukbir J., Blanchard A., 1992, A\&A 262, L21
\bibitem{} Pen U.L., 1998, ApJ 498, 60
\bibitem{} Pierpaoli E., Scott D., White M., 2001, MNRAS 325, 77
\bibitem{} Ponman T.J., Cannon D.B., Navarro F.J., 1999, Nature 397, 135
\bibitem{} Press W., Schecter P., 1974, ApJ 187, 425
\bibitem{}  Reichart D. E., Castander F. J., Nichol R. C., 1999, ApJ 516, 1
\bibitem{} Reichart D. E., Nichol R. C., Castander F. J., Burke D. J., Romer A. K., Holden B. P.,
Collins C. A., Ulmer M. P., 1999, ApJ 518, 521
\bibitem{} Rosati P., Della Ceca R., Norman C., Giacconi R., 1998, ApJ 492, L21
\bibitem{} Rosati P., Borgani S., \& Norman C., 2002, ARAA 40, 539
\bibitem{} Sadat R., Blanchard A., Oukbir J., 1998 A\&A 329, 21
\bibitem{} Sheth R. K., Tormen G., 1999, MNRAS 308, 119
\bibitem{} Sheth R. K., Mo H. J., Tormen G., 2001, MNRAS 323, 1 (SMT)
\bibitem{} Sheth R. K., Tormen G., 2002, MNRAS 329, 61 (ST)
\bibitem{} Tanaka Y., Inoue H., Holt S.S., 1994, PASJ 46, L37
\bibitem{} Tozzi P., Governato F., 1998, "The Young Universe: Galaxy Formation and Evolution at Intermediate and High Redshift". Edited by S. D'Odorico, A. Fontana, and E. Giallongo. ASP Conference Series; Vol. 146; 1998, p.461
\bibitem{} Viana P.T.P, Liddle A.R., 1996 MNRAS 281, 323
\bibitem{} Viana P.T.P., Liddle A.R., 1999, MNRAS 303, 535
\bibitem{} Voit C. M., Donahue M., 1998, ApJ 500, 111 (astro-ph/9804306)
\bibitem{} Voit C. M., 2000, ApJ 543, 113 (astro-ph/0006366)
\bibitem{} Wang L., Steinhardt P.J., 1998, ApJ 508, 483
\bibitem{} Xu H., Jing G., Wu X., 2001, ApJ 553, 78, (astro-ph/0101564)
\end{thebibliography}
\end{document}